\documentclass[10pt,journal,compsoc]{IEEEtran}

\usepackage{amsthm}
\usepackage{times,amsmath,epsfig,amssymb}
\usepackage{graphicx}
\usepackage{balance}  
\usepackage{type1cm}     
\usepackage{algorithm}     
\usepackage{algorithmic}     
\usepackage{xspace}     
\usepackage{booktabs}     
\usepackage[bf,tableposition=top]{caption}     
\usepackage{siunitx}          
\usepackage[hyphens]{url}     
\usepackage[bookmarks=true,bookmarksnumbered=true,
pdfpagemode={UseOutlines},plainpages=false,pdfpagelabels=true,
colorlinks=true,linkcolor={black},citecolor={black},urlcolor={black},
pdftitle={Partial Key Grouping},
pdfsubject={Distributed Systems, Stream Mining},
pdfauthor={Muhammad Anis Uddin Nasir, Gianmarco De Francisci Morales, David Garcia-Soriano, Nicolas Kourtellis, Marco Serafini},
pdfkeywords={Load balancing, stream processing}]{hyperref}     
\usepackage[show]{chato-notes}
\usepackage{soul}
\usepackage[square,numbers]{natbib}     
\DeclareGraphicsExtensions{.pdf,.png,.jpg,.eps}
\graphicspath{{./img/}}

\newcommand{\spara}[1]{\smallskip\noindent\textbf{#1}}

\newenvironment {squishlist}
{\begin{list}{$\bullet$}
  { \setlength{\itemsep}{1pt}
     \setlength{\parsep}{1pt}
     \setlength{\topsep}{1pt}
     \setlength{\partopsep}{1pt}
     \setlength{\leftmargin}{1.5em}
     \setlength{\labelwidth}{1em}
     \setlength{\labelsep}{0.5em} } }
{\end{list}}

\newcommand{\tuple}[1]{\ensuremath{\langle #1 \rangle}\xspace}
\newcommand{\hash}[1]{\ensuremath{\mathcal{H}_{#1}}\xspace}
\newcommand{\pei}{\textsc{pei}\xspace}
\newcommand{\peis}{{\pei}s\xspace}
\newcommand{\pe}{\textsc{pe}\xspace}
\newcommand{\pes}{{\pe}s\xspace}
\newcommand{\potc}{\textsc{p\textup{o}tc}\xspace}
\newcommand{\dspe}{\textsc{dspe}\xspace}
\newcommand{\dspes}{{\dspe}s\xspace}
\newcommand{\dagr}{\textsc{dag}\xspace}

\newcommand{\pkg}{\textsc{Partial Key Grouping}\xspace}
\newcommand{\pkgs}{\textsc{pkg}\xspace}
\newcommand{\kg}{\textsc{kg}\xspace}
\newcommand{\sg}{\textsc{sg}\xspace}
\newcommand{\sources}{\ensuremath{\mathcal{S}}\xspace}
\newcommand{\numsources}{S\xspace}
\newcommand{\workers}{\ensuremath{\mathcal{W}}\xspace}
\newcommand{\numworkers}{W\xspace}
\newcommand{\keyspace}{\ensuremath{\mathcal{K}}\xspace}
\newcommand{\keysize}{K\xspace}
\newcommand{\mycomment}[1]{}

\DeclareMathOperator*{\expect}{\mathbb{E}}
\DeclareMathOperator*{\avg}{avg}
\DeclareMathOperator*{\argmin}{\arg\min}
\DeclareMathOperator*{\argmax}{\arg\max}

\newtheorem{theorem}{Theorem}[section]
\newtheorem{lemma}[theorem]{Lemma}
\newtheorem{corollary}[theorem]{Corollary}

\begin{document}
%
\title{\pkg: Load-Balanced Partitioning of~Distributed~Streams}
%
%
%
%

\author{Muhammad Anis Uddin Nasir,
        Gianmarco De Francisci Morales,\\
        David Garc\'ia-Soriano,
        Nicolas Kourtellis,
        and~Marco Serafini
\IEEEcompsocitemizethanks{\IEEEcompsocthanksitem M. A. Uddin Nasir is with KTH Royal Institute of Technology, Stockholm, Sweden.
E-mail: anisu@kth.se
\IEEEcompsocthanksitem G. De Francisci Morales is with Aalto University, Helsinki, Finland.\protect\\
E-mail: gdfm@acm.org
\IEEEcompsocthanksitem D. Garc\'ia-Soriano and N. Kourtellis are with Yahoo Labs, Barcelona, Spain.
E-mails: davidgs@yahoo-inc.com, nkourtellis@gmail.com
\IEEEcompsocthanksitem M. Serafini is with Qatar Computing Research Institute, Doha, Qatar.\protect\\
E-mail: mserafini@qf.org.qa}
\thanks{Manuscript received XXXXX; revised XXXXXXX.}}

%
%

\markboth{IEEE Transactions on Parallel and Distributed Systems, XXXX}%
{Uddin Nasir \MakeLowercase{\textit{et al.}}: The Power of Both Choices}
%



\IEEEtitleabstractindextext{%
\begin{abstract}
We study the problem of load balancing in distributed stream processing engines, which is exacerbated in the presence of skew.
We introduce \pkg (\pkgs), a new stream partitioning scheme that adapts the classical ``power of two choices'' to a distributed streaming setting by leveraging two novel techniques: \emph{key splitting} and \emph{local load estimation}.
In so doing, it achieves better load balancing than key grouping while being more scalable than shuffle grouping.


We test \pkgs on several large datasets, both real-world and synthetic.
Compared to standard hashing, \pkgs reduces the load imbalance by up to several orders of magnitude, and often achieves nearly-perfect load balance.
This result translates into an improvement of up to 175\% in throughput and up to 45\% in latency when deployed on a real Storm cluster.
\pkg has been integrated in Apache Storm v0.10. 
\end{abstract}

}

\maketitle

\IEEEdisplaynontitleabstractindextext

%
\IEEEpeerreviewmaketitle


%
%
%
%
\IEEEraisesectionheading{\section{Introduction}\label{sec:intro}}

\IEEEPARstart{D}{istributed} stream processing engines (\dspes) such as S4,\footnote{\url{https://incubator.apache.org/s4}} Storm,\footnote{\url{https://storm.apache.org}} and Samza\footnote{\url{https://samza.apache.org}} have recently gained much attention owing to their ability to process huge volumes of data with very low latency on clusters of commodity hardware.
Streaming applications are represented by directed acyclic graphs (\dagr) where vertices, called \emph{processing elements} (\pes), represent operators, and edges, called \emph{streams}, represent the data flow from one \pe to the next.
For scalability, streams are partitioned into sub-streams and processed in parallel on a replica of the \pe called \emph{processing element instance} (\pei).

\begin{figure}[t]
\begin{center}
\includegraphics[scale=0.3]{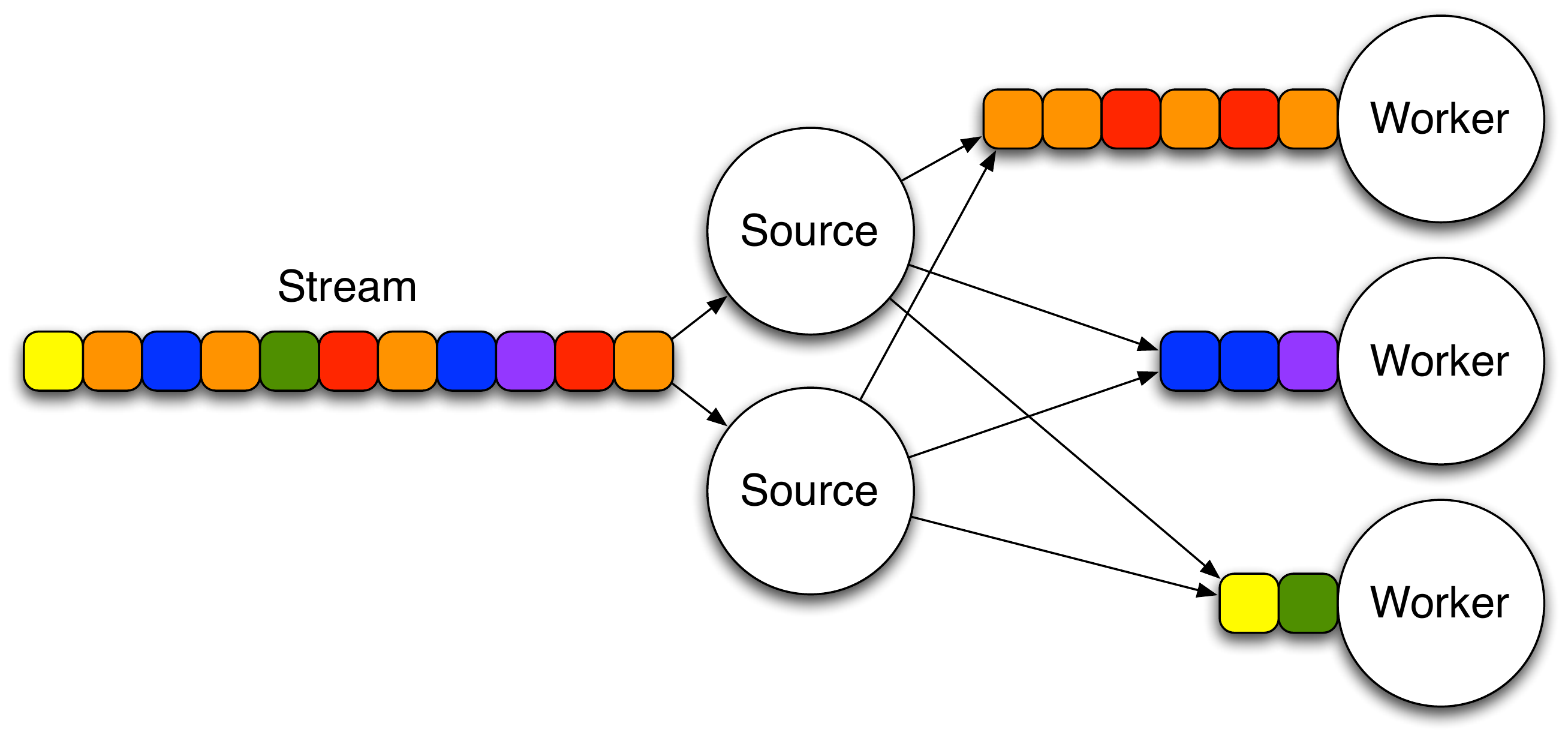}
\caption{Load imbalance generated by skew in the key distribution when partitioning the stream via key grouping. 
The color of each message represents its key.}
\label{fig:imbalance}
\end{center}
\vspace{-\baselineskip}
\end{figure}

Applications of \dspes, especially in data mining and machine learning, usually require accumulating state across the stream by grouping the data on common fields \cite{ben-haim2010spdt,berinde2010heavyhitters}.
Akin to MapReduce, this grouping in \dspes is commonly implemented by partitioning the stream on a \emph{key} and ensuring that messages with the same key are processed by the same \pei.
This partitioning scheme is called \emph{key grouping}, and typically it maps keys to sub-streams by using a hash function.
Hash-based routing allows source \peis to route each message solely via its key,  without needing to keep any state or to coordinate among \peis.
Alas, it also results in high load imbalance as it represents a ``single-choice'' paradigm~\citep{one_choice_load}, and because it disregards the popularity of a key, i.e., the number of messages with the same key in the stream, as depicted in Figure~\ref{fig:imbalance}.

Large web companies run massive deployments of \dspes in production.
Given their scale, good utilization of resources is critical.
However, the skewed distribution of many workloads causes a few \peis to sustain higher load than others. 
This suboptimal load balancing leads to poor resource utilization and inefficiency.


Another partitioning scheme called \emph{shuffle grouping} achieves excellent load balancing by using a round-robin routing, i.e., by sending a message to the next \pei in cyclic order, irrespective of its key.
However, this scheme is mostly suited for stateless computations.
Shuffle grouping may require an additional aggregation phase and more memory to express stateful computations (Section~\ref{sec:preliminaries}).
Additionally, it may cause a decrease in accuracy for some data mining algorithms (Section~\ref{sec:applications}).

This work focuses on the problem of load balancing stateful applications in \dspes when the input stream presents a skewed key distribution.
In this setting, load balancing is attained by having upstream \peis create a balanced partition of messages for  downstream \peis.
Any practical solution for this task needs to be both \emph{streaming} and \emph{distributed}: the former constraint enforces the use of an online algorithm, as the distribution of keys is not known in advance, while the latter calls for a decentralized solution with minimal coordination overhead in order to ensure scalability.


To address this problem, we leverage the ``power of two choices'' (\potc), whereby the system picks the least loaded out of two candidate \peis for each key~\citep{mitzenmacher2001power}.
However, to maintain the semantic of key grouping while using \potc (i.e., so that one key is handled by a single \pei), sources need to track which of the two possible choices has been made for each key.
This requirement imposes a coordination overhead every time a new key appears, so that all sources agree on the choice.
In addition, sources should store this choice in a routing table.
The system needs a table for each source in a stream,  
each with one entry per key.
Given that a stream may contain billions of keys, this solution is not practical.

Instead, we relax the semantic of key grouping, and allow each key to be handled by \emph{both} candidate \peis.
We call this technique \emph{key splitting}.
It allows to apply \potc without the need to agree on, or keep track of, the choices made.
As shown in Section~\ref{sec:evaluation}, key splitting provides good load balance even in the presence of skew.

A second issue is how to estimate the load of a downstream \pei.
Traditional work on \potc assumes global knowledge of the current load of each server, which is challenging in a distributed system.
Additionally, it assumes that all messages originate from a single source, whereas messages in a \dspe are generated in parallel by multiple sources.

In this paper we prove that, interestingly, a simple \emph{local load estimation} technique, whereby each source independently tracks the load of downstream \peis, performs very well in practice.
This technique gives results that are almost indistinguishable from those given by a global load oracle.

The combination of these two techniques (key splitting and local load estimation) enables a new stream partitioning scheme named \emph{\pkg}~\citep{uddin2015pobc}.


In summary, we make the following contributions:
\begin{squishlist}
\item We study the problem of load balancing in modern distributed stream processing engines.
\item We propose \pkg, a novel and simple stream partitioning scheme that applies to any \dspe.
When implemented on top of Apache Storm, it requires a single function and less than 20 lines of code.\footnote{Available at \url{https://github.com/gdfm/partial-key-grouping}}
\item \pkg shows how to apply \potc to \dspes in a principled and practical way, and we propose two novel techniques to do so: key splitting and local load estimation.
\item We measure the impact of \pkg on a real deployment on Apache Storm.
Compared to key grouping, it improves the throughput of an example application on real-world datasets by up to $175\%$, and the latency by up to $45\%$.
\item Our technique has been integrated into Apache Storm v0.10, and is available in its standard distribution.\footnote{\url{https://issues.apache.org/jira/browse/STORM-632}}
\end{squishlist}

\section{Preliminaries and Motivation}
\label{sec:preliminaries}



We consider a \dspe running on a cluster of machines that communicate by exchanging messages over the network by following the flow of a \dagr, as discussed.
In this work, we focus on balancing the data transmission along a single edge in a \dagr.
Load balancing across the whole \dagr is achieved by balancing each edge independently.
Each edge represents a single stream of data, along with its partitioning scheme.
Given a stream under consideration, let the set of upstream \peis (sources) be \sources, and the set of downstream \peis (workers) be \workers, and their sizes be $|\sources| = \numsources$ and $|\workers| = \numworkers$ (see Figure~\ref{fig:imbalance}).

The input to the engine is a sequence of messages $m = \tuple{t,k,v}$ where $t$ is the timestamp at which the message is received, $k \in \keyspace \,, |\keyspace| = \keysize$ is the message's key, and $v$ its value.
The messages are presented to the engine in ascending order of timestamp.

A \emph{stream partitioning} function $P_{t}: \mathcal{K} \rightarrow \mathbb{N}$ maps each key in the key space to a natural number, at a given time $t$.
This number identifies the worker responsible for processing the message.
Each worker is associated to one or more keys.

We use a definition of \emph{load} similar to others in the literature (e.g., Flux~\citep{shah2003flux}).
At time $t$, the load of a worker $i$ is the number of messages handled by the worker up to $t$:
$$ L_i(t) = |\{ \tuple{\tau,k,v} : P_{\tau}(k) = i \wedge \tau \leq t \}|. $$

In principle, depending on the application, two different messages might impose a different load on workers.
However, in most cases these differences even out and modeling such application-specific differences is not necessary.

We define {\em imbalance} at time $t$ as the difference between the maximum and the average load of the workers:
$$ I(t) = \max_{i}(L_i(t)) - \avg_{i}(L_i(t)), \mathrm{\ for\ } i \in \workers. $$

We tackle the problem of identifying a stream partitioning function that minimizes the imbalance,
while at the same time avoiding the downsides of shuffle grouping, highlighted next.

%


\subsection{Existing Stream Partitioning Functions}
\label{sec:existing-partitioning}
Several primitives are offered by \dspes to partition the stream, i.e., for sources to route outgoing messages to different workers.
There are two main primitives of interest: \emph{key grouping} (\kg) and \emph{shuffle grouping} (\sg).

\kg ensures that messages with the same key are handled by the same \pei (analogous to MapReduce).
It is usually implemented through hashing. 
\sg routes messages independently, typically in a round-robin fashion.
\sg provides excellent load balance by assigning an almost equal number of messages to each \pei.
However, no guarantee is made on the partitioning of the key space, as each occurrence of a key can be assigned to any \peis.
\sg is the perfect choice for \emph{stateless} operators.
When running \emph{stateful} operators, one has to handle, store, and aggregate multiple partial results for the same key, thus incurring additional costs. 


In general, when the distribution of input keys is skewed, the number of messages that each \pei needs to handle can be very different.
While this problem is not present for stateless operators, which can use \sg to evenly distribute messages, stateful operators implemented via \kg suffer from load imbalance.
This issue generates a degradation of the service level, or reduces the utilization of the cluster which must be provisioned to handle the peak load of the single most loaded server.


\spara{Example.}
To make the discussion more concrete, we introduce a simple application that will be our running example: the \emph{na\"{i}ve Bayes classifier}.
A na\"{i}ve Bayes classifier is a probabilistic model that assumes independence of features in the data (hence the na\"{i}ve).
It estimates the probability of a class $C$ given a feature vector $X$ by using Bayes' theorem:
$$
P(C | X) = \frac{ P(X | C) P(C) }{ P(X) }.
$$
The answer given by the classifier is then the class with maximum likelihood
$$
C^* = \argmax_C P(C | X).
$$
Given that features are assumed independent, the joint probability of the features is the product of the probability of each feature. Also, we are only interested in the class that maximizes the likelihood, so we can omit $P(X)$ from the maximization as it is constant.
The class probability is proportional to the product
$$
P(C | X) \propto \prod_{x_i \in X} P(x_i | C) P(C),
$$
which reduces the problem to estimating the probability of each feature value $x_i$ given a class $C$, and a prior for each class $C$.

In practice, the classifier estimates the probabilities by counting the frequency of co-occurrence of each feature and class value.
Therefore, it can be implemented by a set of counters, one for each pair of feature value and class value.
A MapReduce implementation is straightforward, and available in Apache Mahout.\footnote{\url{https://mahout.apache.org/users/classification/bayesian.html}}

\spara{Implementation via key grouping.} Following the MapReduce paradigm, the implementation of na\"{i}ve Bayes uses \kg on the source stream.
Let us assume we want to train a classifier from a stream of documents.
Each document is split into its constituent words by a tokenizer \pe.
The tokenizer also adds the class to each word.
By keying on the word, the data is sent to a counter \pe, which keeps a running counter for each word-class pair.
\kg ensures that each word is handled by a single \pei, which thus has the total count for the word in the stream.

\begin{figure}[t]
\begin{center}
\includegraphics[width=\columnwidth]{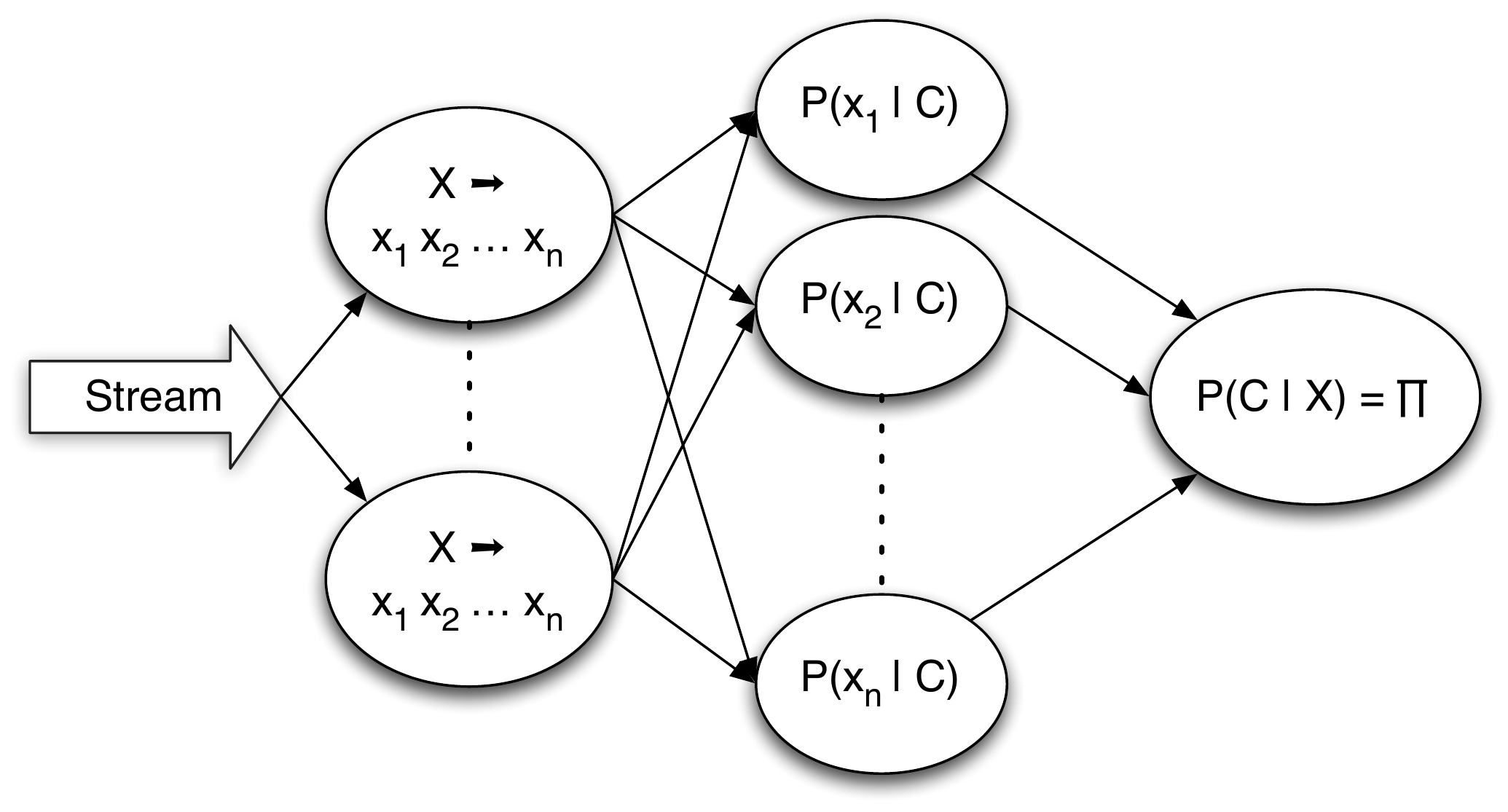}
\caption{Na\"{i}ve Bayes implemented via key grouping (\kg).}
\label{fig:vnb}
\end{center}
\vspace{-\baselineskip}
\end{figure}

When we want to classify a document, we split it into its constituent words, and send each word to the counter \pei responsible for it.
Each \pei will return the probability for each word-class pair, which can be combined by class by a downstream \pe.
The aggregation can use \kg on a transaction ID (e.g., the document ID) to gather all the probabilities.
This process just multiplies the probabilities for each class (more typically sums them given that we keep the log-likelihood), and returns the maximum as the predicted class (see Figure~\ref{fig:vnb}).

Clearly, the use of \kg generates load imbalance as, for instance, the \pei associated to the key ``the" will receive many more messages than the one associated with ``Barcelona".
This example captures the core of the problem we tackle: the distribution of word frequencies follows a Zipf law, where few words are extremely common while a large majority are rare.
Therefore, an even distribution of keys, such as the one generated by \kg, results in an uneven distribution of messages.

\spara{Implementation via shuffle grouping.}
An alternative implementation uses shuffle grouping on the source stream to get several partial models.
Each model is trained on an independent sub-stream of the original document stream.
These models are sent downstream to an aggregator every $T$ seconds via key grouping.
The aggregator simply combines the counts for each key to get the total count, and thus the final model.
This implementation resembles the use of combiners in MapReduce, where each mapper generate a partial result before sending it to the reducers.

An alternative implementation simply keeps the model partitioned across the workers, and queries all of them in parallel when a document needs to be classified.
This implementation trades off aggregation at training time for increased query latency at classification time, which is usually contrary to the goal of a streaming classifier.
Therefore, we describe this implementation only for completeness, and consider only the former one henceforth.

Using \sg requires a slightly more complex logic but it generates an even distribution of messages among the counter \peis.
However, it suffers from other problems.
Given that there is no guarantee on which \pei will handle a key, each \pei potentially needs to keep a counter for \emph{every} key in the stream.
Therefore, the memory usage of the application grows linearly with the parallelism level.
Hence, it is not possible to scale to a larger workload by adding more machines: the application is not scalable in terms of memory.
Even if we resort to approximation algorithms, in general, the error depends on the number of aggregations performed, thus it grows linearly with the parallelism level.
We analyze this case in further detail along with other application scenarios in Section~\ref{sec:applications}.

%

\subsection{Key grouping with rebalancing}
One common solution for load balancing in \dspes is \pei migration~\citep{shah2003flux,cherniack2003scalable,xing2005dynamic,gedik2013partitioning,balkesen2013adaptive,castro2013integrating}.
Once a situation of load imbalance is detected, the system activates a rebalancing routine that moves part of the \peis, and the state associated with them, away from an overloaded worker.
While this solution is easy to understand, its application in our context is not straightforward. 


Rebalancing requires setting a number of parameters such as how often to check for imbalance and how often to rebalance.
These parameters are often application-specific as they involve a trade-off between imbalance and rebalancing cost that depends on the size of the state to migrate. 

Further, implementing a rebalancing mechanism usually requires major modifications of the \dspe at hand.
This task may be hard, and is usually seen with suspicion by the community driving open source projects, as witnessed by the many variants of Hadoop that were never merged back into the main line of development~\citep{abouzeid2009hadoopdb,Dittrich2010hadooppp,yang2007mapreducemerge}.

In our context, rebalancing implies migrating keys from one sub-stream to another.
However, this migration is not directly supported by the programming abstractions of some \dspes.
Storm and Samza use a \emph{coarse-grained} stream partitioning paradigm.
Each stream is partitioned into as many sub-streams as the number of downstream \peis.
Key migration is not compatible with this partitioning paradigm, as a key cannot be uncoupled from its sub-stream. 
In contrast, S4 employs a \emph{fine-grained} paradigm where the stream is partitioned into one sub-stream per key value, and there is a one-to-one mapping of a key to a \pei.
The latter paradigm easily supports migration, as each key is processed independently.


A major problem with mapping keys to \peis explicitly is that the \dspe must maintain several routing tables: one for each stream. 
Each routing table has one entry for each key in the stream.
Keeping these tables is impractical because the memory requirements are staggering.
In a typical web mining application, each routing table can easily have billions of keys.
For a moderately large \dagr with tens of edges, each with tens of sources, the memory overhead easily becomes prohibitive.

Finally, as already mentioned, for each stream there are several sources sending messages in parallel.
Modifications to the routing table must be consistent across all sources, so they require coordination, which creates further overhead.
For these reasons we consider an alternative approach to load balancing. 

\section{Partial Key Grouping}
\label{sec:algos}


The problem described so far currently lacks a satisfying solution.
To solve this issue, we resort to a widely-used technique in the literature of load balancing: the so-called \emph{``power of two choices''} (\potc).
While this technique is well-known and has been analyzed thoroughly both from a theoretical and practical perspective~\citep{adler1995parallelrandomized,azar1999balanced-allocations,byers2003geometricgeneralizations,lenzen2011parallelrandomized,mitzenmacher2001power,mitzenmacher2001potc-survey}, its application in the context of \dspes is not straightforward and has not been previously studied.


Introduced by \citet{azar1999balanced-allocations}, \potc is a simple and elegant technique that allows to achieve load balance when assigning units of load to workers.
It is best described in terms of ``balls and bins''.
Imagine a process where a stream of balls (units of work) is distributed to a set of bins (the workers) as evenly as possible.
The \emph{single-choice paradigm} corresponds to putting each ball into one bin selected uniformly at random.
By contrast, the power of two choices selects two bins uniformly at random, and puts the ball into the least loaded one.
This simple modification of the algorithm has powerful implications that are well known in the literature (see Sections~\ref{sec:theory},~\ref{sec:rel-work}). 


\spara{Single choice.}
The current solution used by all \dspes to partition a stream with key grouping corresponds to the single-choice paradigm.
The system has access to a single hash function $\hash{1}(k)$.
The partitioning of keys into sub-streams is determined by the function $ P_{t}(k) = \hash{1}(k) \bmod W $.

The single-choice paradigm is attractive because of its simplicity: the routing does not require to maintain any state and can be done independently in parallel.
However, it suffers from a problem of load imbalance~\citep{mitzenmacher2001power}.
This problem is exacerbated when the distribution of input keys is skewed.

\spara{PoTC.}
With the power of two choices, the system has two hash functions $\hash{1}(k)$ and $\hash{2}(k)$.
The algorithm maps each key to the sub-stream assigned to the least loaded between the two possible workers:
$P_{t}(k)$~=~$\argmin_i(L_i(t) : \hash{1}(k) = i \lor \hash{2}(k) = i)$.

The theoretical gain in load balance with two choices is exponential compared to a single choice.
However, using more than two choices only brings constant factor improvements~\citep{azar1999balanced-allocations}.
Therefore, we restrict our study to two choices.

\potc introduces two additional complications.
First, to maintain the semantics of key grouping, the system needs to \emph{keep state} and track the choices made.
Second, the system has to \emph{know the load} of the workers in order to make the right choice.
We discuss these two issues next.






\subsection{Key Splitting}

A na\"{i}ve application of \potc to key grouping requires the system to store a bit of information for each key seen, to keep track of which of the two choices needs to be used thereafter.
This variant is referred to as \emph{static} \potc henceforward.

Static \potc incurs some of the same problems discussed for key grouping with rebalancing.
Since the actual worker to which a key is routed is determined dynamically, sources need to keep a routing table with an entry per key.
As already discussed, maintaining this routing table is often impractical.

In order to leverage \potc and make it viable for \dspes, we relax the requirement of key grouping.
Rather than mapping each key to one of the two possible choices, we allow it to be mapped to \emph{both choices}.
Every time a source sends a message, it selects the worker with the lowest current load among the two candidates associated to that key.
This technique, called \emph{key splitting}, introduces several new trade-offs.

First, key splitting allows the system to operate in a decentralized manner, by allowing multiple sources to take decisions independently in parallel.
As in key grouping and shuffle grouping, no state needs to be kept by the system and each message can be routed independently.

Second, key splitting enables far better load balancing compared to key grouping.
It allows using \potc to balance the load on the workers:
by splitting each key on multiple workers, it handles the skew in the key popularity.
Moreover, given that \emph{all} its decisions are dynamic and based on the current load of the system (as opposed to static \potc), key splitting adapts to changes in the popularity of keys over time.

Third, key splitting reduces memory usage and aggregation overhead compared to shuffle grouping.
Given that each key is assigned to \emph{exactly two} \peis, the memory to store its state is only a constant factor higher than when using \kg.
Instead, with \sg the memory grows linearly with the number of workers $W$.
Additionally, state aggregation needs to happen only once for the two partial states, as opposed to $W - 1$ times in shuffle grouping.
This improvement also allows to reduce the error incurred during aggregation for some algorithms, as discussed in Section~\ref{sec:applications}.


For an application developer, key splitting gives rise to  a novel stream partitioning scheme called \pkg, which lies in-between key grouping and shuffle grouping.
 
Naturally, not all algorithms can be expressed via \pkgs.
The functions that can leverage \pkgs are the same ones that can leverage a combiner in MapReduce, i.e., associative functions and monoids.
Examples of applications include na\"{i}ve Bayes, heavy hitters, and streaming parallel decision trees, as detailed in Section~\ref{sec:applications}.
On the contrary, other functions such as computing the median cannot be easily expressed via \pkgs.


\spara{Example.}
Let us examine the streaming na\"{i}ve Bayes example using \pkgs.
In this case, each word is tracked by two counters on two different \peis. 
Each counter holds a partial count for the word-class pairs, while the total count is the sum of the two partial counts.
Therefore, the total memory usage is $2 \times K$, i.e., $O(K)$.
Compare this result to \sg where the memory is $O(WK)$.
Partial counts are sent downstream via \kg to an aggregator that computes the final model.
For each word-class pair, the application sends two counters, and the aggregator performs a constant time aggregation.
The total work for the aggregation is $O(K)$.
Conversely, with \sg the total work is again $O(WK)$.
Compared to the implementation with \kg, the one with \pkgs requires additional logic, some more memory and has some aggregation overhead.
However, it also provides a much better load balance which maximizes the resource utilization of the cluster.
The experiments in Section~\ref{sec:evaluation} prove that the benefits outweigh its cost.




\subsection{Local Load Estimation}
\potc requires knowledge of the load of each worker to take its decisions.
A \dspe is a distributed system, and, in general, sources and workers are deployed on different machines.
Therefore, the load of each worker is not readily available to each source.


Interestingly, we prove that no communication between sources and workers is needed to effectively apply \potc.
We propose a \emph{local load estimation} technique, whereby each source independently maintains a local load-estimate vector with one element per worker.
The load estimates are updated by using only local information of the portion of stream sent by each source.
We argue that in order to achieve global load balance it is sufficient that each source independently balances the load it generates across all workers.

The correctness of local load estimation directly follows from our standard definition of load in Section~\ref{sec:preliminaries}.
The load on a worker $i$ at time $t$ is simply the sum of the loads that each source $j$ imposes on the given worker:
$ L_i(t) = \sum_{j \in \sources}{L_i^j(t)}. $
Each source $j$ can keep an estimate of the load on each worker $i$ based on the load it has generated $L_i^j$. 
As long as each source keeps its own portion of load balanced, then the overall load on the workers will also be balanced.
Indeed, the maximum overall load is at most the sum of the maximum load that each source sees locally. 
It follows that the maximum imbalance is also at most the sum of the local imbalances. 
Therefore, we can bound the overall imbalance by a function of local imbalances:
\begin{eqnarray*}
I(t) = \max_i(L_i(t)) - \avg_i(L_i(t)) & =  &\\
\max_i(\sum_{j}{L_i^j(t)}) - \avg_i(\sum_{j}{L_i^j(t)}) 
&\leq &\\
\sum_{j}{\max_i(L_i^j(t))} - \sum_{j}{\avg_i(L_i^j(t))} & =  &  \sum_{j}{\hat{I}^j(t)} \\
\end{eqnarray*}
where $\hat{I}^j(t)$ is the local imbalance estimated at source $j$.
Consequently, by minimizing $\hat{I}^j(t)$ for each source we also minimize the upper bound on the actual imbalance.


\section{Applications}
\label{sec:applications}

\pkgs is a novel programming primitive for stream partitioning and not every algorithm can be expressed with it.
In general, all algorithms that use shuffle grouping can use \pkgs to reduce their memory footprint.
In addition, many algorithms expressed via key grouping can be rewritten to use \pkgs in order to get better load balancing.
In this section we provide a few such examples of common data mining algorithms, and show the advantages of \pkgs. 
Henceforth, we assume that each message contains a data point for the application, e.g., a feature vector in a high-dimensional space.

\subsection{Streaming Parallel Decision Tree}


A decision tree is a classification algorithm that uses a tree-like model where nodes are tests on features, branches are possible outcomes, and leafs are class assignments.

\citet{ben-haim2010spdt} propose an algorithm to build a streaming parallel decision tree that uses approximated histograms to find the test value for continuous features.
Messages are shuffled among $W$ workers.
Each worker generates histograms independently for its sub-stream, one histogram for each feature-class-leaf triplet.
These histograms are then periodically sent to a single aggregator that merges them to get an approximated histogram for the whole stream.
The aggregator uses this final histogram to grow the model by taking split decisions for the current leaves in the tree.
Overall, the algorithm keeps $W \times D \times C \times L$ histograms, where $D$ is the number of features, $C$ is the number of classes, and $L$ is the current number of leaves.

The memory footprint of the algorithm depends on $W$, so it is impossible to fit larger models by increasing the parallelism.
Moreover, the aggregator needs to merge $W \times D \times C$ histograms each time a split decision is tried, and merging the histograms is one of the most expensive operations.

Instead, \pkgs reduces both the space complexity and aggregation cost.
If applied on the features of each message, a single feature is tracked by two workers, with an overall cost of only $2$$\times$$D$$\times$$C$$\times$$L$ histograms.
Furthermore, the aggregator needs to merge only two histograms per feature-class-leaf triplet.
This scheme allows to alleviate memory pressure by adding more workers, as the space complexity does not depend on $W$.


\subsection{Heavy Hitters and Space Saving}

The heavy hitters problem consists in finding the top-k most frequent items occurring in a stream.
The \textsc{SpaceSaving}~\citep{metwally2005spacesaving} algorithm solves this problem approximately in constant time and space.
Recently, \citet{berinde2010heavyhitters} have shown that \textsc{SpaceSaving} is space-optimal, and how to extend its guarantees to merged summaries.
This result allows for parallelized execution by merging partial summaries built independently on separate sub-streams.

In this case, the error bound on the frequency of a single item depends on a term representing the error due to the merging, plus another term which is the sum of the errors of each individual summary for a given item $i$:
$$
\mid \hat{f}_i - f_i \mid \leq \Delta_{f} + \sum_{j=i}^{W}\Delta_{j}
$$
where $f_i$ is the true frequency of item $i$ and $\hat{f}_i$ is the estimated one, each $\Delta_j$ is the error from summarizing each sub-stream, while $\Delta_{f}$ is the error from summarizing the whole stream, i.e., from merging the summaries.

Observe that the error bound depends on the parallelism level $W$.
Conversely, by using \kg, the error for an item depends only on a single summary, thus it is equivalent to the sequential case, at the expense of poor load balancing.
Using \pkgs we achieve both benefits: the load is balanced among workers, and the error for each item depends on the sum of only two error terms, regardless of the parallelism level.

%

\section{Analysis}
\label{sec:theory}

\newcommand{\naturals}{{\mathbb N}}
\DeclareRobustCommand{\calA}[0]{{\mathcal A}}
\DeclareRobustCommand{\calP}[0]{{\mathcal P}}
\DeclareRobustCommand{\calQ}[0]{{\mathcal Q}}
\DeclareRobustCommand{\calD}[0]{{\mathcal D}}
\DeclareRobustCommand{\calS}[0]{{\workers}}
\DeclareRobustCommand{\calK}[0]{{\keyspace}}
\newcommand{\code}[1]{{\textsc #1}}
\newcommand{\indic}{\mathbb{I}\,}


We proceed to analyze the conditions under which \pkgs achieves good load balance.
Recall from Section~\ref{sec:preliminaries} that we have a set $\calS$ of $n$ workers at our disposal and receive a sequence of $m$ messages  $k_1,
       \ldots, k_m$ with values from a key universe $\calK$. Upon receiving the $i$-th message with value $k_i\in \calK$, we need to decide its placement 
among the workers; decisions are irrevocable. We assume one message arrives per unit of time.
Our goal is to minimize the eventual maximum load $L(m)$, which is the same as minimizing the imbalance $I(m)$.
A simple placement scheme such as shuffle grouping provides an imbalance of at most one, but
we would like to limit the number of workers processing each key to $d \in \naturals^+$.

\spara{Chromatic balls and bins.}
We model our problem in the framework of balls and bins processes, where keys correspond to colors, messages to colored balls, and workers to bins.
Choose
 $d$ independent hash
functions $\hash{1}, \ldots, \hash{d}\colon \calK \to [n]$ uniformly at random. 
Define the \code{greedy-$d$} scheme as follows:
at time $t$, the $t$-th ball (whose color is $k_t$) is placed on the bin with minimum current load among
$\hash{1}(k_t), \ldots, \hash{d}(k_t)$, i.e., $P_t(k_t) = \operatorname{argmin}_{i \in \{\hash{1}(k_t),\ldots,\hash{d}(k_t)\}} L_i(t)$.
Recall that with key splitting there is no need to remember the choice for the next time a ball of the same color appears.

Observe that when $d = 1$,
each ball color is assigned to a unique bin so no choice has to be made; this models hash-based key grouping.
At the other extreme, when $d \gg n \ln n$, all $n$  bins are valid choices, and we obtain shuffle
grouping.

\spara{Key distribution.}
Finally, we assume an underlying discrete distribution $\calD$ supported on $\calK$ from which ball colors
are drawn, i.e., $k_1, \ldots, k_m$ is a sequence of $m$ independent samples from $\calD$. 
Without loss of generality, we identify the set $\calK$ of keys with
$\naturals^+$ or, if $\calK$ is finite with cardinality $\keysize = |\calK|$, where $[\keysize] = \{1, \ldots, \keysize\}$.
We assume them ordered by decreasing probability: if $p_i$ is the probability of drawing key $i$ from $\calD$, then
$p_1 \ge p_2 \ge p_3 \ge \ldots \ge 0$ and $\sum_{i \in \calK} p_i = 1$. We also identify the set $\calS$ of bins with~$[n]$.

\subsection{Imbalance with \pkg}

\spara{Comparison with standard problems.}
As long as we keep getting balls of different colors, our process is identical to the standard \code{greedy-$d$} process of \citet{azar1999balanced-allocations}.
This occurs with high probability provided that 
$m$ is small enough.
But for sufficiently large $m$ (e.g., when $m \ge \frac{1}{p_1}$), repeated keys will start to arrive.
Recall that for any number of choices $d\ge 2$, the maximum imbalance after throwing $m$ balls \emph{of different colors} into $n$ bins with the standard \code{greedy-$d$} process is $\frac{\ln \ln n}{\ln d} + \frac{m}{n} + O(1)$. 
Unfortunately, such strong bounds (independent of $m$) cannot apply to our setting.
To gain some intuition on what may go wrong, consider the following examples where $d$$=$$2$.

Note that for the maximum load not to be much larger than the average load, the number of bins used must not exceed $O(1/p_1)$, where $p_1$ is the maximum key probability.
Indeed, at any time we expect
the two bins $h_1(1), h_2(1)$ to contain together at
least a $p_1$ fraction of all balls, just counting the occurrences of a single key. Hence the expected maximum load among the two grows at a rate
of at least $p_1/2$ per unit of
time, while the overall average load increases by exactly $\frac{1}{n}$ per unit of time. Thus, if $p_1 > 2/ n$, the expected imbalance at time $m$ will be lower bounded by
$(\frac{p_1}2 - \frac1n) m$, which grows \emph{linearly} with $m$. This holds irrespective of the placement scheme used.


However, requiring $p_1 \le 2 / n$ is not enough to prevent imbalance $\Omega(m)$. Consider the uniform distribution
over $n$ keys. Let $B = \bigcup_{i \le n} \{ \hash{1}(i), \hash{2}(i) \}$ be the set of all bins that belong to one of the potential choices for some key. 
An easy application of linearity of expectation shows that the expected size of $B$ is $n - n \left(1 - \frac{1}{n}\right)^{2n} \approx n (1 -
    \frac{1}{e^2})$. So all $n$ keys use only an $(1-\frac{1}{e^2}) \approx 0.865$ fraction of all
bins, and roughly $0.135 n$ bins will remain unused. In fact the imbalance after $m$ balls
will be at least $\frac{m}{0.865 n}-\frac{m}{n} \approx 0.156 m$.
The problem is that most concrete instantiations of our two random hash functions cause the existence of
an ``overpopulated'' set $B$ of bins inside which
the average bin load must grow faster than the average load across all bins.
(In fact, this case subsumes our first example above, where $B$ was $\{ \hash{1}(1), \hash{2}(1) \}$.)
%

Finally, even in the absence of overpopulated bin subsets, some inherent imbalance
is     due to deviations between the empirical and true key distributions.
For instance, suppose there are two
keys $1, 2$ with equal probability $\frac12$ and $n =4$ bins. With constant probability, key 1 is assigned to bins $1, 2$
and key 2 to bins $3, 4$. This situation looks perfect because the \code{greedy-2} choice will send each
occurrence of key 1 to bins $1,2$ alternately so the loads of bins $1, 2$ will always equal up to $\pm 1$.
However, the number of balls with key 1 seen is likely to deviate from $m/2$ by roughly $\Theta(\sqrt m)$, so either the top two or the bottom two bins will receive
$m / 4 + \Omega(\sqrt m)$ balls, and the imbalance will be $\Omega(\sqrt m)$.

In the remainder of this section we carry out our analysis, which broadly construed asserts that the above are the only impediments to achieve good balance. 

\spara{Statement of results.}
We  noted  that once the number of bins exceeds $2/p_1$ (where $p_1$ is the maximum key
frequency), the maximum load will be dominated by the loads of the bins to which the most frequent key is mapped. 
Hence the main case of interest is where $p_1 = O(\frac{1}{n})$.


We focus on the case where the number of balls is large compared to the number of bins. 
The following results show that partial key grouping can significantly reduce the maximum load (and the imbalance), compared to key
grouping.
\begin{theorem}\label{thm:main}
Suppose we use $n$ bins and let $m \ge n^2$.
Assume a key distribution $\calD$ with maximum probability $p_1 \le \frac{1}{5 n}$.
Then
    the imbalance after $m$ steps of the \code{Greedy-$d$} process satisfies, with probability at least $1-\frac{1}{n}$,
$$ I(m) = 
\begin{cases}
O\left(\frac{m}{n} \cdot \frac{\ln n}{\ln \ln n}\right), & \text{if } d = 1\\
O\left(\frac{m}{n}\right), & \text{if } d \geq 2\\
    \end{cases}
    . $$
\mycomment{
the following hold:
\begin{itemize}
\item For any $d\ge 2$, the imbalance after $m$ steps of the \code{Greedy-$d$} process satisfies, with probability at least $1-\frac{1}{n}$,
$$ I(m) = \Theta\left(\frac{m}{n}\right) . $$
Moreover, with probability at least $1-\frac{1}{n}$ over the choice of the $d$ hash functions, any load balancing policy using the same set of
choices for each key must have imbalance
$\Omega\left(\frac{m}{n}\right)$.
\item For some distributions $\calD$,
the imbalance after $m$ steps of the \code{Greedy-1} process satisfies, with probability at least $1-\frac{1}{n}$,
$$ I(m) = \Theta\left(\frac{m}{n} \cdot \frac{\ln n}{\ln \ln n}\right) . $$
\end{itemize}
}
\end{theorem}

%

These bounds are best-possible:\footnote{However, the imbalance can be much smaller than the worst-case bounds from
        Theorem~\ref{thm:main} if the probability of most keys is much smaller than $p_1$, which is the case in many setups.}
    there is a distribution $\calD$ satisfying the hypothesis of Theorem~\ref{thm:main} such that
    the imbalance after $m$ steps of the \code{Greedy-$d$} process satisfies, with probability at least $1-\frac{1}{n}$,
    $$ I(m) = 
    \begin{cases}
    \Omega\left(\frac{m}{n} \cdot \frac{\ln n}{\ln \ln n}\right), & \text{if } d = 1\\
    \Omega\left(\frac{m}{n}\right), & \text{if } d \geq 2\\
        \end{cases}
        . $$

    In fact, this is the case when $\calD$ is the     
uniform        distribution over a set of $5 n$ keys (the proof is straightforward and hence omitted).

    The next section is devoted to the proof of the upper bound, Theorem~\ref{thm:main}. 

\subsection{Proof}
\mycomment{
First define $\tau_d^\calD$.

\begin{theorem}
Conditioned on $\tau_d^\calD$, with high probability after $m$ throws the normalized imbalance after running the \code{Greedy-$d$} process satisfies
$$
\widehat{I}(m) \le (\tau_d^\calD - 1) + \sqrt{\frac{n \ln n \ln m}{m}}.
$$
Also, the optimal normalized imbalance after the hash functions are chosen satisfies
$$\widehat{I}^*(m) \ge (\tau_d^\calD - 1) + \sqrt{\frac{\ln n}{m}}.$$
\end{theorem}

\begin{theorem}
With high probability,
     $$\tau_d^\calD = \min(O(\frac{\ln n}{\ln \ln n}), \frac{1}{n} + \sqrt{p_1 n \ln n}$$
and
        for $d \ge 2$,
        $$\tau_d^\calD = \min(O(\frac{1}{n} + p_1)).$$
\end{theorem}

In particular, when $p_1 \le \frac{1}{n^3 \ln n}$, the maximum load will be $O(m / n)$ whether we use one choice or $d > 1$ choices.
On the other hand,
From these results we see that when $p_1 \ge \frac{1}{n}$, the imbalance with $d = 1$ or
In the typical case we consider where $p_1 < \frac{1}{n}$,

}

\spara{The $\mu_r$ measure of bin subsets.}
For every nonempty set of bins $B \subseteq[n]$ and $1 \le r \le d$, define
$$ \mu_r(B) = \sum_i\{ p_i \mid \{\hash{1}(i), \ldots, \hash{r}(i)\} \subseteq B \} .$$
We will be  interested in $\mu_1(B)$ (which measures the probability that a random key from $\calD$ will have its choice inside $B$) and $\mu_d(B)$
(which measures the probability that a random key from $\calD$ will have all its choices inside $B$).
Note that $\mu_1(B) = \sum_{j \in B} \mu_1(\{j\})$ and $\mu_d(B) \le \mu_{d-1}(B)$ for $d > 1$.

A key component of the proof will be to  show that, for small enough bin subsets ($B \subseteq [n]$ where $|B| \le \frac{n}{5}$), it holds that
$\mu_d(B) \le \frac{|B|}{n}$. The intuition for the usefulness of this property is the following. Let~$B$ denote the set of bins that are likely to be
``highly overloaded'' (according to some suitable definition). Assuming that we can argue separately that the load of all bins outside $B$ is smaller
than the maximum load in $B$, it follows that the probability that a random key from 
$\mathcal D$ increases the load of some bin in $B$ is at most $\frac{|B|}{n}$, which is no worse than the probability of the same event were we to
use \code{Greedy-$n$} instead of~\code{Greedy-$d$}.
This will enable us to conclude that the load imbalance caused by \code{Greedy-$d$} is also small.

\spara{Connection to expander graphs.}
To understand why such a property must hold, it helps to restate it in graph-theoretical terms. 
Construct a bipartite graph $G$ with keys on the left, bins on the right, and edges from keys to their bin choices.
For every key subset $A \subseteq \mathcal K$, define a weight $p(A) = \sum_i p_i$ and 
let $\Gamma(A) = \bigcup_i \{\hash{1}(i), \ldots, \hash{r}(i)\} \}$ denote the neighbourhood of~$A$. Then the property
   ``$\mu_d(B) \le \frac{|B|}{n}$ whenever $|B| \le \frac{n}{5}$'' is equivalent
to ``$|\Gamma(A)| \ge \min(n\cdot p(A), n/5)$ for all $A$''.
Now it becomes clear that our claim amounts to stating that $G$ is a kind of \emph{vertex expander graph}~\cite{pseudorandomness,expander_graphs}.
For example, suppose for simplicity that we have $n$ bins and $5n$ keys. Then the property we seek then says that the neighborhood of every set of $t
\le n$
vertices on the left side has size at least $t / 5$.
It is well known that left-regular bipartite random graphs enjoy these vertex-expansion properties, and our claim may be viewed as a generalization of these facts to certain node-weighted graphs (where the weight of a left vertex $i$ is given by $p_i$). 

\spara{Concentration inequalities.}
We recall the following results (see \cite{concentration} for a reference), which we need to prove our main theorem.
\begin{theorem}[Chernoff bounds]\label{chernoff}
Suppose $\{X_i\}$ is a finite sequence of independent random variables with $X_i \in [0, M]$ and let $Y = \sum_i X_i$, $\mu = \sum_i \expect[X_i]$.
Then, for all $\delta \ge 0$,
$$ \Pr[ Y \ge (1 + \delta) \mu ] \le \left( \frac{e^{\delta}}{(1+\delta)^{1+\delta}} \right)^{\frac{\mu}{M}} .$$
Therefore, for all $\beta \ge \mu$,
$$ \Pr[ Y \ge \beta ] \le C(\mu, \beta, M) , $$
where
$$ C(\mu, \beta, M) \triangleq \exp\Big(-\frac{\beta \ln(\frac{\beta}{e \mu}) + \mu}{M}\Big). $$
\end{theorem}

\begin{theorem}[McDiarmid's inequality]\label{bounded_dif}
Let $X_1, \ldots, X_n$ be a vector of independent random variables and let $f$ be a real-valued function satisfying
    $ |f(a) - f(a') | \le 1$
whenever the vectors $a$ and $a'$ differ in just one coordinate. Then, for all $\lambda \ge 0$,
$$ \Pr[  f(X_1, \ldots, X_n) > \expect[ f(X_1, \ldots, X_n) ] + \lambda ] \le \exp(-2 \lambda^2).$$
\end{theorem}

\begin{lemma}\label{lem:mu1}
For every $B \subseteq [n]$, it holds that
     $\expect [\mu_1(B)]= \frac{|B|}{n}$. Moreover, if $p_1 \le \frac{1}{n}$, for any $\lambda > 0$ it holds that
       $$ \Pr \left[ \mu_1(B) \ge \frac{|B|}{n} (e \lambda) \right]  \le \left(\frac{1}{\lambda^\lambda}\right)^{|B|}. $$
\end{lemma}
\begin{proof}
The first claim follows from linearity of expectation and the fact that $\sum_i p_i = 1$:
\begin{align*}
\expect[\mu_1(B)] &= \expect\left[ \sum_i  p_i \cdot \indic[ \hash{1}(i) \in B] \right]  \\
                  &= \sum_i p_i \Pr[ \hash{1}(i) \in B ] = \sum_i p_i \frac{|B|}{n} = \frac{|B|}{n}. \\
\end{align*}                  
For the second, let $|B| = k$. Using Theorem~\ref{chernoff} with $X_i = p_i \cdot \indic[ \hash{1}(i) \in B ] \in [0, p_1]$,
    we obtain that
$\Pr \left[ \mu_1(B) \ge \frac{k}{n} (e \lambda) \right]$ is at most
    $$
C\left(\frac{k}{n}, \frac{k}{n} e \lambda, p_1\right)
                                                   \le \exp\left(-\frac{k}{n p_1} e \lambda \ln \lambda \right) 
                                                   \le \exp(-k \lambda \ln \lambda),$$

since $n p_1 \le 1$.
\end{proof}

\mycomment{
    \begin{corollary}
    With high probability, for all $B \subseteq [n]$ it holds that
    $$ \mu_1(B) \le O\left(\frac{|B|}{n} \frac{\ln n}{\ln \ln n}\right). $$
    \end{corollary}
    \begin{proof}
    Since $\mu_1(B) = \sum_{j \in B} \mu_1(\{j\})$, it suffices to show the claim when $|B| = 1$.
    This follows from Lemma~\ref{lem:mu1} by setting $|B| = 1$, $\lambda = O(\frac{\ln n}{\ln \ln n})$, and applying the union bound.
    \end{proof}
}

\begin{lemma}\label{lem:mu2}
For every $B \subseteq [n]$,
     $\expect [\mu_d(B)]= \left(\frac{|B|}{n}\right)^d$
and, provided that $p_1 \le \frac{1}{5 n}$,
       $$ \Pr \left[ \mu_d(B) \ge \frac{|B|}{n} \right]  \le \left(\frac{e |B|}{n}\right)^{5 |B|}. $$
\end{lemma}
\begin{proof}
Again, the first claim is straightforward. For the second, let $|B| = k$. Using Theorem~\ref{chernoff},
$\Pr \left[ \mu_d(B) \ge \frac{k}{n} \right]$ is at most
\begin{align*}
 C\left(\Big(\frac{k}{n}\Big)^d, \frac{k}{n}, p_1\right)
                                            &\le \exp\left(-\frac{k (d - 1)}{n p_1} \ln \left(\frac{n}{e k}\right) \right) \\
                                            &\le \exp\left(-5 k \ln \left(\frac{n}{e k}\right)\right)
\end{align*}    
    
since $n p_1 \le \frac{1}{5}$.
\end{proof}
\begin{corollary}\label{coro:mu2}
Assume $p_1 \le \frac{1}{4n}$, $d \ge 2$.  Then, with high probability, 
       $$ \max \left\{ \frac{\mu_d(B)}{|B| / n} \middle| B \subseteq [n], |B| \le \frac{n}{5} \right\} \le 1.$$
\end{corollary}
\begin{proof}
We use Lemma~\ref{lem:mu1} and the union bound. The probability that the claim fails to hold is bounded by
\begin{align*}
 \sum_{\substack{B \subseteq [n]\\|B| \le n / 5}} \Pr \left[ \mu_d(B) \ge \frac{k}{n} \right] &\le \sum_{k=1}^{n/5} \binom{n}{k} \left(\frac{e k}{n}\right)^{5k} \\
  \le \sum_{k=1}^{n/5} \left(\frac{e n}{k}\right)^k \left(\frac{e k}{n}\right)^{5k} &= \sum_{k=1}^{n/5} \left(\frac{e^{3/2} k}{n}\right)^{4k}                                                                    = o\left(\frac{1}{n}\right),
\end{align*}                                                                  
where we use  $\binom{n}{k} \le \left(\frac{e n}{k}\right)^k$, valid for all $k$.
\end{proof}

For a scheduling algorithm $\calA$ and a set $B \subseteq [n]$ of bins, write $L^\calA_B(t) = 
\max_{j\in B} L_j(t)$ for the maximum load among the bins in $B$ after $t$ balls have been
processed by $\calA$.
\begin{lemma}\label{lem:perfect}
Suppose there is a set $A \subseteq [n]$ of bins such that for all $T \subseteq A$,
        $ \mu_d(T) \le \frac{|T|}{n}. $
Then $\calA=\text{\code{Greedy-$d$}}$ satisfies $L^\calA_A(m) = O(\frac{m}{n}) + L^\calA_{[n]\setminus A}(m)$ with high probability.
\end{lemma}
\begin{proof}
We use a coupling argument~\cite{concentration}. Consider the following two independent processes $\calP$ and $\calQ$: $\calP$ proceeds as
\code{Greedy-$d$}, while $\calQ$ picks the bin for each ball independently at
random from $[n]$ and increases its load. Consider any time~$t$ at which the load vector is $\omega_t \in \naturals^n$
and $M_t=M(\omega_t)$ is the set of bins with
maximum load. After handling the $t$-th ball, let $X_t$ denote the event that $\calP$ increases the maximum load in $A$ because the new ball has all choices in
$M_t \cap A$, and $Y_t$ denote the event that $\calQ$ increases the maximum load in $A$. Finally, let $Z_t$ denote the event
that $\calP$ increases the maximum load in $A$ because the new ball has some choice in $M_t \cap A$ and some choice in $M_t \setminus
A$, but the load of one of  its choices in $M_t \cap A$ is no larger. We identify these events with their indicator
random variables.

Note that the maximum load in $A$ at the end of Process $\calP$ is $L_A^\calP(m) = \sum_{t\in[m]} (X_t + Z_t)$, while  at the
end of Process $\calQ$ is $L_A^\calQ(m) = \sum_{t\in[m]}
Y_t$. Conditioned on any load vector $\omega_t$, the probability of $X_t$ is $$\Pr[ X_t \mid \omega_t] =
\mu_d(M_t \cap A) \le \frac{|M_t \cap A|}{n}  \le \frac{|M_t|}{n} = \Pr[ Y_t \mid \omega_t ],$$
    So $\Pr [ X_t   \mid
\omega_t ] \le \Pr [Y_t \mid \omega_t]$, which implies that for any $b \in \naturals$, $\Pr [ \sum_{t \in [m]} X_t
\le b ] \ge \Pr[
\sum_{t \in [m]} Y_t
\le b ].$ But with high probability, the maximum load of Process $\calQ$ is $b = O(m / n)$, so $\sum_t X_t = O(m / n)$
holds with at least the same probability. On the other hand, $\sum_t Z_t \le L_{[n]\setminus A}^\calP(m)$ because each occurrence of
$Z_t$ increases the maximum load in $A$, and once a time $t$ is reached such that $L_A^\calP(t) > L_{[n]\setminus
    A}^\calP(m)$, event $Z_t$ must cease to happen. Therefore
$L_A^\calP(m) = \sum_{t\in[m]} X_t + \sum_{t\in[m]} Z_t \le O(m / n) + L_{[n]\setminus A}^\calP(m)$,
yielding the result.
\end{proof}

\begin{proof}[Proof of Theorem~\ref{thm:main}]
Let $$A = \left\{j \in [n] \mid \mu_1(\{j\}) \ge \frac{3 e}{n}\right\}.$$
Observe that every bin $j \notin A$ has $\mu_1(\{j\}) < \frac{3 e}{n}$. Assume for the moment that we used the \code{Greedy-1} process that simply throws every ball to the \emph{first}
    choice; then, by the Chernoff bound, the probability that $j \notin A$ and the eventual load of bin $j$ after $m \ge n^2$ throws exceeds $20 m / n > 2 (3e m /n)$ is exponentially small in $n$.
Therefore, in this situation, the maximum load of all bins not in $A$ is at most $20 / n$ with high probability.
The same result holds for \code{Greedy-$d$} because of the majorization technique of
Azar et al~\cite[Theorem 3.5]{azar1999balanced-allocations}. Therefore our task reduces to showing that the maximum load of the bins in $A$ is $O(\frac{m}{n})$.

Consider the sequence $X_1, \ldots, X_\keysize$ of random variables given by $X_i = \hash{1}(i)$, and let $f(X_1, X_2,
        \ldots, X_\keysize) = |A|$ denote the number of bins $j$ with $\mu_1(\{j\}) \ge \frac{3 e}{n}$.
By Lemma~\ref{lem:mu1}, 
$ \expect[|A|] = \expect[ f ] = \sum_{i\in[n]} \Pr[\mu_1({i}) \ge 3e / n] \le \frac{n}{27} $.
Moreover, the function~$f$
satisfies the hypothesis of Theorem~\ref{bounded_dif}: a change in the random choice of $\hash{1}{i}$ may only affect the size of $|A|$ by one.
We conclude that, with high probability, $|A| \le \frac{n}{5}.$

Now assume that the thesis of Corollary~\ref{coro:mu2} holds, which happens except with probability $o(1/n)$.
Then we have that for all $B \subseteq A$, $\mu_d(B) \le \frac{|B|}{n}.$
Thus, Lemma~\ref{lem:perfect} applies to $A$. This means that after throwing $m$ balls, the maximum load among the bins in $A$ is $O(\frac{m}{n})$, as
we wished to show.
\end{proof}

\section{Evaluation}
\label{sec:evaluation}

We assess the performance of our proposal by using both simulations and real deployments.
In so doing, we answer the following questions:
\begin{itemize}
\item[\textbf{Q1:}]
What is the effect of key splitting on \potc?
\item[\textbf{Q2:}]
How does local estimation compare to a global oracle?
\item[\textbf{Q3:}]
How robust is \pkg to changes in the skew of the keys? 
\item[\textbf{Q4:}]
What is the effect of number of choices on \pkg's performance?
\item[\textbf{Q5:}]
What is the overall effect of \pkg on applications deployed on a real \dspe?
\end{itemize}


\subsection{Experimental Setup}

\begin{table}[t]
\caption{Summary of the datasets used in the experiments: number of messages, number of keys and percentage of messages having the most frequent key ($p_1$).} 
\centering
\small
\begin{tabular}{l c c c r}
\toprule
Dataset		&	Symbol	&	Messages		&	Keys			& 	$p_1$(\%)		\\
\midrule
Wikipedia		&	WP		&	\num{22}M	&	\num{2,9}M	&	\num{9,32}	\\
Twitter		&	TW		&	\num{1,2}G	&	\num{31}M	&	\num{2,67}	\\
Cashtags		&	CT		&	\num{690}k	&	\num{2,9}k	&	\num{3,29}	\\
\midrule
LiveJournal	&	LJ		&	\num{69}M	&	\num{4,9}M	&	\num{0,29}	\\
Slashdot0811	&	SL$_1$	&	\num{905}k	&	\num{77}k		&	\num{3,28}	\\
Slashdot0902 	&	SL$_2$	&	\num{948}k	&	\num{82}k		&	\num{3,11}	\\
\midrule
Lognormal 1	&	LN$_1$	&	\num{10}M	&	\num{16}k		&	\num{14,71}	\\
Lognormal 2	&	LN$_2$	&	\num{10}M	&	\num{1,1}k	&	\num{7,01}	\\
Zipf			&	ZF		&    \num{10}M   	&  \num{1}k,\dots,\num{1}M &   $\propto$ $\frac{1} { \sum{x^{-z}} }$ \\
\bottomrule
\end{tabular}
\label{tab:summary-datasets}
\end{table}


\spara{Datasets.}
Table~\ref{tab:summary-datasets} summarizes the datasets used.
We use two main real datasets, one from \emph{Wikipedia} and one from \emph{Twitter}.
These datasets were chosen for their large size, different degree of skewness, and different set of applications in Web and online social network domains.
The Wikipedia dataset (WP)\footnote{\url{http://www.wikibench.eu/?page\_id=60}} is a log of the pages visited during a day in January 2008. 
Each visit is a message and the page's URL represents its key.
The Twitter dataset (TW) is a sample of tweets crawled during July 2012.
Each tweet is split into words, which are used as the key for the message.

\begin{figure}[t]
\begin{center}
	\includegraphics[width=\columnwidth]{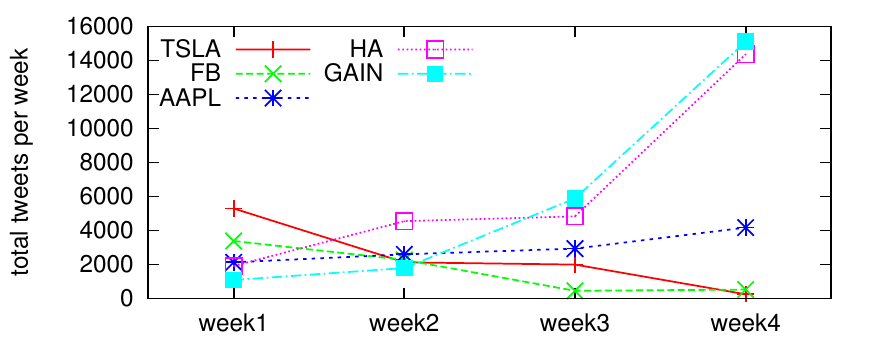}
	\caption{Frequency of tweets for the top 5 tickers in CT.
	The most frequent keys change throughout time.}
	\label{fig:key-frequency-distribution-tickers}
\end{center}
\vspace{-\baselineskip}
\end{figure}

Additionally, we use a Twitter dataset that comprises of tweets crawled in November 2013.
The keys for the messages are the \emph{cashtags} in these tweets.
A \emph{cashtag} is a ticker symbol used in the stock market to identify a publicly traded company preceded by the dollar sign (e.g., \$AAPL for Apple).
As shown in Figure~\ref{fig:key-frequency-distribution-tickers}, popular cash tags change from week to week.
This dataset allows to study the effect of shift of skew in the key distribution.

Moreover, we experiment on three additional datasets comprised of directed graphs\footnote{\url{http://snap.stanford.edu/data}} (LJ, SL$_1$, SL$_2$).
We use the edges in the graph as messages and the vertices as keys. 
These datasets are used to test the robustness of \pkgs to skew in partitioning the stream \emph{at the sources}, as explained next.
They also represent a different kind of application domain: streaming graph mining.

Furthermore, we generate two synthetic datasets with keys following a log-normal distribution (LN$_1$, LN$_2$), a commonly used heavy-tailed skewed distribution~\citep{talwar2007weightedcase}.
The parameters of the distribution ($\mu_1$=$1.789$, $\sigma_1$=$2.366$; $\mu_2$=$2.245$, $\sigma_2$=$1.133$) come from an analysis of Orkut, and emulate workloads from the online social network domain~\cite{benevenuto09characterizingusers}.
Lastly, we generate synthetic datasets with keys following Zipf distributions with exponent in the range $z = \{ 0.1, \ldots, 2.0 \}$ and for different number of unique keys $K$~=~$1$k$, 10$k$, 100$k$,$ and $1$M.
Each unique key of rank \emph{r} appears with frequency $f$ as follows:
$$f(r,K,z) = \frac {1/r^z }{\sum_{x=1}^{K} (1/x^z)}.$$

\spara{Simulation.}
We process the datasets by simulating the \dagr presented in Figure~\ref{fig:imbalance}, which represents the simples possible topology.
The stream is composed of timestamped keys that are read by multiple independent sources (\sources) via shuffle grouping, unless otherwise specified.
The sources forward the received keys to the workers (\workers) downstream.
In our simulations we assume that the sources perform data extraction and transformation, while the workers perform data aggregation, which is the most computationally expensive part of the \dagr.
Thus, the workers are the bottleneck in the \dagr and the focus for the load balancing.




\subsection{Experimental Results}



\begin{figure*}[t]
\begin{center}
	\includegraphics[width=\textwidth]{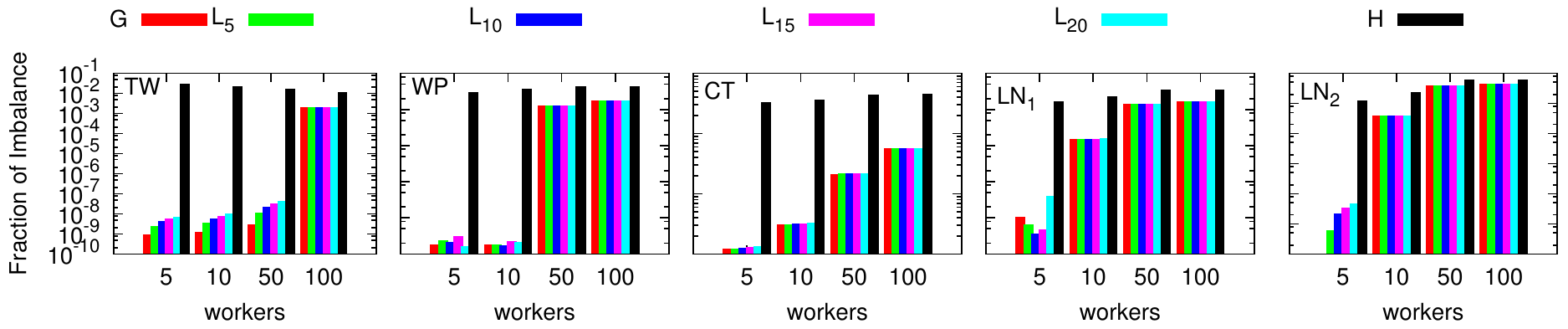}
	\caption{Fraction of average imbalance with respect to total number of messages for each dataset, for different number of workers and number of sources.
	The imbalance of local estimation is very close to the global oracle, and is not affected by the number of sources.} 
\label{fig:avg-GLH}
\end{center}
\vspace{-\baselineskip}
\end{figure*}

\spara{Q1.}
We measure the imbalance in the simulations when using the following techniques:
\begin{squishlist}
\item[H:] Hashing, which represents standard key grouping (\kg) and is our main baseline.
We use a $64$-bit Murmur hash function to minimize the probability of collision.
\item[\potc:] Power of two choices \emph{without} using key splitting, i.e., traditional \potc applied to key grouping.
\item[On-Greedy:] Online greedy algorithm that picks the least loaded worker to handle a new key.
\item[Off-Greedy:] Offline greedy sorts the keys by decreasing frequency and executes On-Greedy.
\item[\pkgs:] \potc with key splitting. 
\end{squishlist}

\begin{table}[t]
\caption{Fraction of average imbalance with different numbers of workers for the WP and TW datasets.
\kg causes large imbalance, up to $\approx 9\%$ on WP.
\pkgs performs consistently better than other methods, with negligible imbalance up to 50 workers on TW.}
\tabcolsep=0.1cm
\centering
\footnotesize
\begin{tabular}{l r r r r r r r r}
\toprule
Dataset		&	\multicolumn{4}{c}{WP} 					&	\multicolumn{4}{c}{TW}					\\
\cmidrule(lr){2-5} \cmidrule(lr){6-9}
$W$			&	5		&	10		&	50		&	100		&	5		&	10		&	50		&	100		\\
\midrule
\pkgs		&	3.7e-8	&	1.3e-7	&	2.7e-2	&	3.7e-2	&	3.4e-10	&	1.4e-9	&	2.3e-9	&	3.4e-3	\\
Off-Greedy	&	3.7e-8	&	4.1e-8	&	7.4e-2	&	8.3e-2	&	3.4e-10	&	6e-10	&	6.7e-3 	&	1.7e-2	\\
On-Greedy	&	3.6e-7	&	6.4e-3	&	7.4e-2	&	8.3e-2	&	7.2e-9	&	7.9e-8	&	1.0e-2	&	1.7e-2	\\
\potc			&	7.3e-7	&	7.8e-3	&	7.4e-2	&	8.3e-2	&	1.9e-5	&	4.3e-6	&	1.2e-2	&	1.7e-2	\\
Hashing \scriptsize{(\kg)}		&	6.4e-2	&	7.8e-2	&	9.2e-2	&	9.2e-2	&	3.5e-2	&	3.2e-2	&	2.0e-2	&	2.8e-2	\\
\bottomrule
\end{tabular}
\label{tab:key-splitting}
\end{table}

Note that \pkgs is the only method that uses key splitting.
Off-Greedy knows the whole distribution of keys so it represents an unfair comparison for online algorithms.

Table~\ref{tab:key-splitting} shows the results of the comparison on the two main datasets WP and TW.
Each value is the fraction of average imbalance measured throughout the simulation.
As expected, hashing performs the worst, creating a large imbalance in all cases.
While \potc performs better than hashing in all the experiments, it is outclassed by On-Greedy on TW.
On-Greedy performs very close to Off-Greedy, which is a good result considering that it is an online algorithm.
Interestingly, \pkgs performs even better than Off-Greedy.
Relaxing the constraint of \kg allows to achieve a load balance comparable to offline algorithms.

We conclude that \potc alone is not enough to guarantee good load balance, and key splitting is fundamental not only to make the technique practical in a distributed system, but also to make it effective in a streaming setting.
As expected, increasing the number of workers also increases the average imbalance.
The behavior of the system is binary: either well balanced or largely imbalanced.
The transition between the two states happens when $W$ surpasses the limit $O(1/p_1)$ described in Section~\ref{sec:theory}, which happens around $50$ workers for WP and $100$ for TW.



%

\spara{Q2.}
Given the aforementioned results, we focus our attention on \pkgs henceforth.
So far, it still uses global information about the load of the workers when making the choice.
Next, we experiment with local estimation, i.e., each source performs its own estimation of the worker load, based on its past sub-stream.

We consider the following alternatives:
\begin{itemize}
\item[G:] \pkgs with global information of worker load.
\item[L:] \pkgs with local estimation of worker load and different number of sources, e.g., L$_5$ denotes $\numsources=5$.
\item[LP:] \pkgs with local estimation and \emph{periodic probing} of worker load every $T_p$ minutes. For instance, L$_5$P$_1$ denotes $\numsources=5$ and $T_p=1$.
When probing is executed, the local estimate vector is set to the actual load of the workers.
\end{itemize}

Figure~\ref{fig:avg-GLH} shows the average imbalance (normalized to the size of the dataset) with different techniques, for different number of sources and workers, and for several datasets.
The baseline (H) always imposes very high load imbalance on the workers.
Conversely, \pkgs with local estimation (L) has always a lower imbalance.
Furthermore, the difference from the global variant (G) is always less than one order of magnitude.
Finally, this result is robust to changes in the number of sources.



%

\begin{figure*}[t]
\begin{center}
	\includegraphics[scale=0.74]{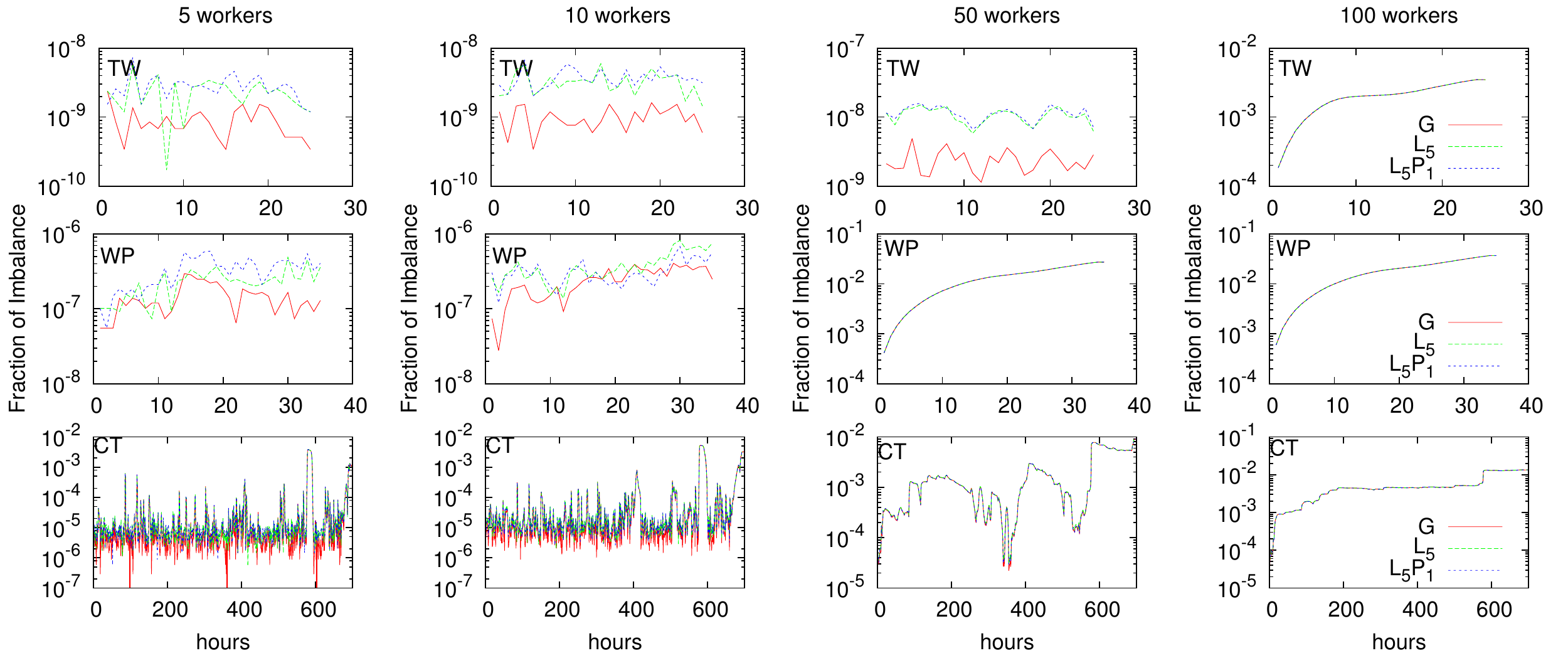}
	\caption{Fraction of average imbalance through time for different datasets, techniques, and number of workers, with $S=5$.
	Probing (L$_5$P$_1$) does not improve the local load estimation (L$_5$), whose performance is anyway always close to the global oracle (G).
	The performance is consistent throughout time, and only depends the number of workers for the given dataset.}
	\label{fig:avg-time}
\vspace{-\baselineskip}
\end{center}
\end{figure*}

Figure~\ref{fig:avg-time} displays the imbalance of the system through time $I(t)$ for TW, WP and CT, 5 sources, and for $W=5, \dots, 100$.
\pkgs has negligible imbalance by using either global information (G) or local estimation (L$_5$).
The only situation where we observe imbalance is when the set of workers is too large for the given set of keys.
In this case, as shown by the analysis in Section~\ref{sec:theory}, each worker will only process a limited number of keys, so the workers responsible for ``hot'' keys will get an unbalanced share of load.

\begin{figure}[t]
\begin{center}
	\includegraphics[scale=1]{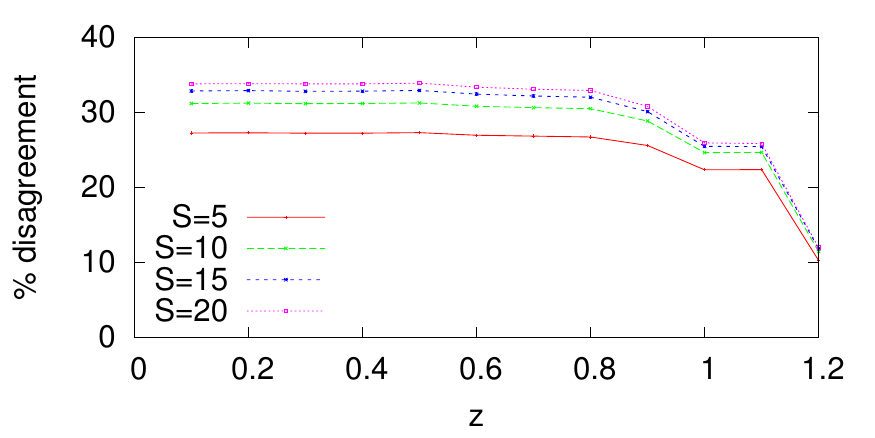}
	\caption{Percentage of disagreement of decisions by the sources with local load estimation in comparison to the global oracle (ZF with $K=10$k and $W=5$). Local load estimation achieves good load balance despite the presence of high disagreement.}
\label{fig:disagreements}
\end{center}
\vspace{-\baselineskip}
\end{figure}

\begin{figure*}[t]
\begin{center}
	\includegraphics[scale=0.74]{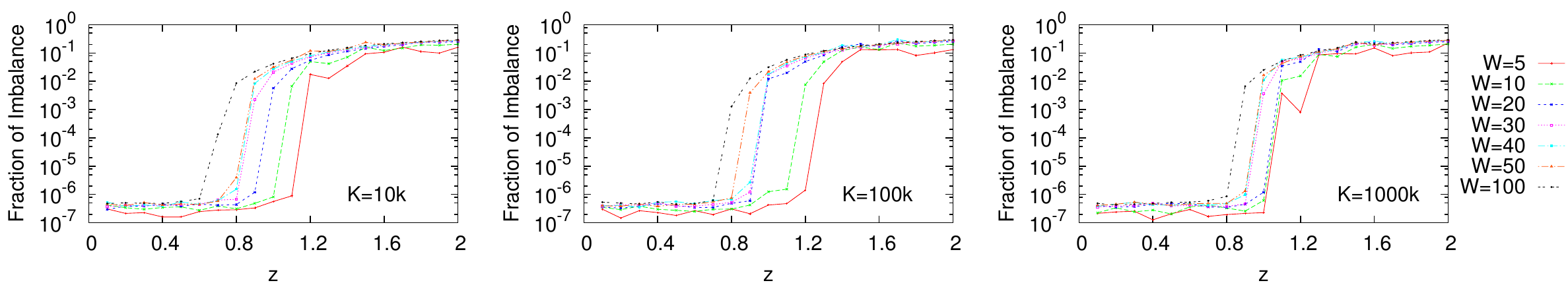}
	\caption{Fraction of average imbalance for \pkgs, while varying the skew of the key distribution, and increasing number of total keys submitted and the number of workers in the system.
	The transition between balanced and unbalanced happens when $p_1$ is too large.}
\label{fig:skews_zipfs}
\end{center}
\vspace{-\baselineskip}
\end{figure*}

Interestingly, even though both G and L achieve very good load balance, their choices are quite different.
In an experiment on the WP dataset, the agreement on the destination of each message between G and L is only 47\% (Jaccard overlap).
We conduct additional experiments with the ZF workload, where key popularity follows a Zipf distribution. 
In the experiments we also increase the number of sources in the system and study the percentage of disagreement in decisions made by the sources in comparison to a global oracle (results shown in Figure~\ref{fig:disagreements}).
Given that we are interested only in the difference in choices between G and L, we limit the experiment in a region of the parameter space where good load balance is attainable, so as to make their choices comparable.
For this to happen, as shown in Figure~\ref{fig:skews_zipfs}, the Zipf exponent $z$ needs to be below $1.2$.

Increasing the skew makes a few keys dominate the distribution.
This skew forces the sources to make practically the same decisions as the oracle, as most keys will be sent to the choice that does not conflict with a frequent key, and therefore the disagreement is reduced.
This is observed regardless of how many sources are present in the system.
These results verify our idea that a good load balance is achievable even when using distributed sources and taking decisions that differ from an oracle (as shown in Figure~\ref{fig:avg-time}).
This fact is true regardless of how large the skew is and how many sources are employed.

Subsequently, L reaches a local minimum in imbalance which is very close in value to the one obtained by G, although via different decisions.
Also, in this case, good balance can only be achieved up to a number of workers that depends on the dataset.
When that number is exceeded, the imbalance increases rapidly, as seen in the cases of WP and partially for CT for $W=50$, where all techniques lead to the same high load imbalance, in accordance to what discussed in Section~\ref{sec:theory}.

Finally, we compare the local estimation strategy with a variant that makes use of periodic probing of workers' load every minute (L$_5$P$_1$).
Probing removes any inconsistency in the load estimates that the sources may have accumulated.
However, interestingly, this technique does not improve the load balance, as shown in Figure~\ref{fig:avg-time}.
Even increasing the frequency of probing does not reduce imbalance (not shown in the figure for clarity).
In conclusion, local information is sufficient to obtain good load balance, therefore it is not necessary to incur the overhead of probing.

\spara{Q3.}
We perform three types of experiments.
First, we examine how robust \pkgs is to an increasing skew in the distribution of keys.
For this experiment, we use the ZF workload.
Figure~\ref{fig:skews_zipfs} shows the fraction of average imbalance when varying the skew of the key distribution.
The experiments show a consistent and stable trend independent of the number of keys.
Thus, we conjecture that the robustness of the approach is not affected by this parameter.
Instead, it highly depends on the number of workers and the skew of the key distribution, as already observed.
In all cases, having an excessive number of workers can lead to imbalance, since the system will not operate at its saturation point.

Note that \pkgs can withstand large skews on the key distribution, up to a threshold level that depends on the specific distribution (for Zipf, $z \approx 1.2$).
Nevertheless, an extremely high skew can still lead to load imbalance.
This issue begs the question of whether using a larger number of choices $d > 2$ might be a solution to this problem, as the bound $p_1 < d/W$ would be satisfied.
We investigate this possibility with the next question.

Second, we use the directed graphs datasets to test the robustness of \pkgs to \emph{skew in the sources}, i.e., when each source forwards an uneven part of the stream.
To do so, we distribute the messages to the sources using \kg.
We simulate a simple application that computes a function of the incoming edges of a vertex (e.g., in-degree, PageRank).
The input keys for the source \pe is the source vertex id, while the key sent to the worker \pe is the destination vertex id, i.e., the source \pe inverts the edge.
This schema projects the out-degree distribution of the graph on sources, and the in-degree distribution on workers, both of which are highly skewed.

\begin{figure}[t]
\begin{center}
	\includegraphics[width=\columnwidth]{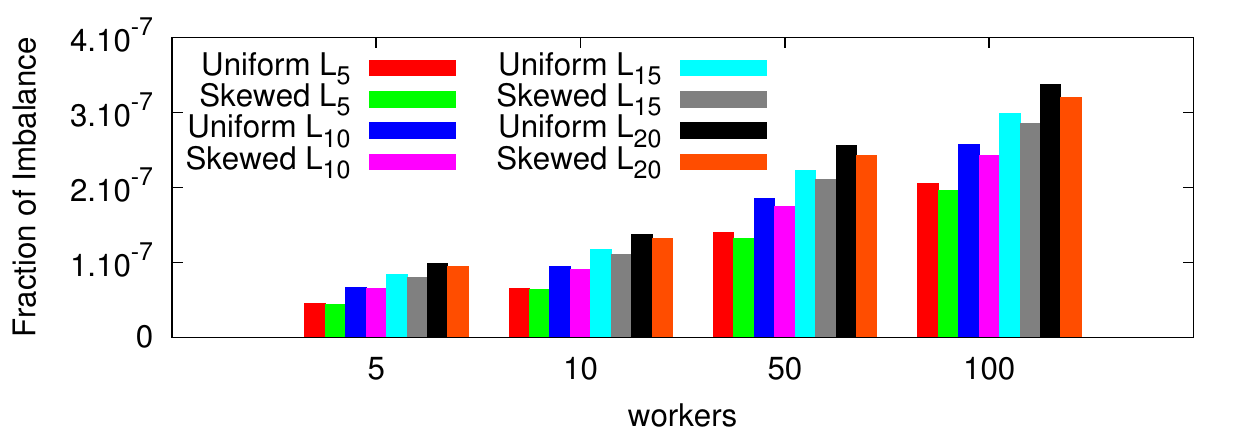}
	\caption{Fraction of average imbalance with uniform and skewed splitting of the input keys on the sources when using the LJ graph.
	Presence of skew at the sources and increase in their number do not affect the performance of local load estimation.}
	\label{fig:avg-skewed-shuffled}
\end{center}
\vspace{-\baselineskip}
\end{figure}

Figure~\ref{fig:avg-skewed-shuffled} shows the average imbalance for the experiments with a skewed split of the keys to sources for the LJ social graph (results on SL$_1$ and SL$_2$ are similar to LJ, omitted due to space constraint).
For comparison, we include the results when the split is performed \emph{uniformly} using shuffle grouping of keys on sources.
On average, the imbalance generated by the skew on sources is similar to the one obtained with uniform splitting.
As expected, the imbalance slightly increases as the number of sources and workers increase, but, 
in general, it remains at very low absolute values.

Third, we experiment with drift in the skew distribution by using the cashtag dataset (CT).
The bottom row of Figure~\ref{fig:avg-time} demonstrates that all techniques achieve a low imbalance, even though the change of key popularity through time generates occasional spikes.

In conclusion, \pkgs is robust to skew on the sources, and can therefore be chained to key grouping.
It is also robust to the drift in key distribution common of many real-world streams.
However, for a large enough skew on the key distribution, \pkgs with two choices can also fail, regardless of the number of available workers.
Next, we investigate if this issue can be resolved with a larger number of choices.

\spara{Q4.}
As shown in Figure~\ref{fig:skews_zipfs} and explained in detail in Section~\ref{sec:theory}, under extreme skew \pkgs may fail to keep imbalance low.
In fact, given a Zipf exponent of $1.2$ in the ZF workload, the system is led to high imbalance regardless of the number of workers.
Therefore, under this setup, we investigate if increasing the number of choices $d$ can improve the imbalance in the system.
While it is well known that increasing $d$ to a number larger than $2$ only generates constant-factor improvements in load balance~\cite{azar1999balanced-allocations}, for practical purposes using a larger $d$ may allow to achieve load balance in configurations for which \pkgs is not sufficient.
The price to pay for this capability is the potential increase in memory usage, from a factor of $2$ to a factor of $d$ higher than with \kg.

\begin{figure}[t]
\begin{center}
	\includegraphics[scale=1]{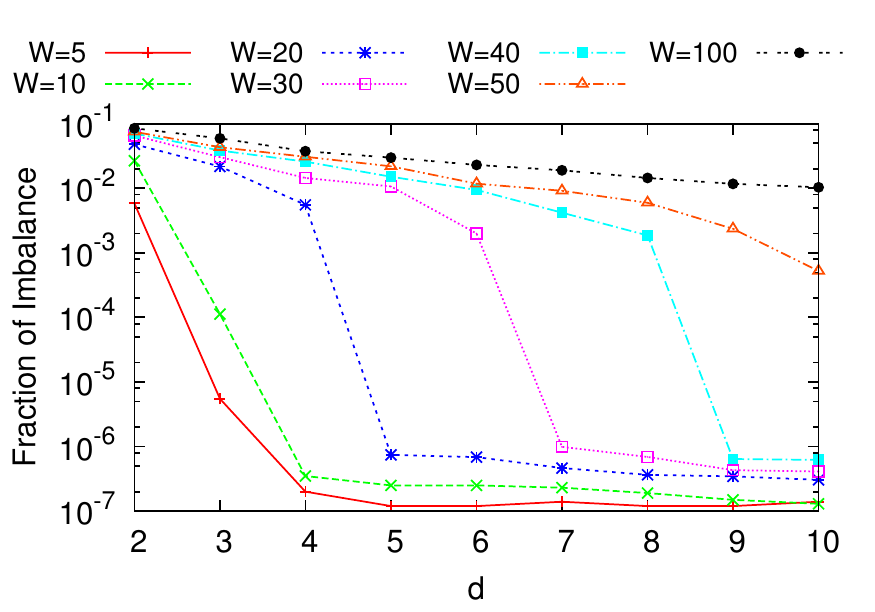}
	\caption{Fraction of average imbalance when varying the number of choices $d$ given to the sources (ZF with $z=1.2$ and $K=1M$). Increasing the number of choices enables good load balance in the presence of skew even with larger numbers of workers.}
\label{fig:dchoices}
\end{center}
\vspace{-\baselineskip}
\end{figure}

Figure~\ref{fig:dchoices} demonstrates that load balance in the system can be restored if the number of choices is increased: from two to four when the workers are five, or to nine when the workers are forty.
For even larger number of workers (e.g., $W=50$ or $100$), the imbalance is still high but can be lowered with a few tens of choices.
The memory cost for this configuration is still lower than the upper bound $O(WK)$ given by \sg, where all workers can potentially receive all available keys.

\begin{figure*}[t]
\begin{center}
	\includegraphics[scale=0.78]{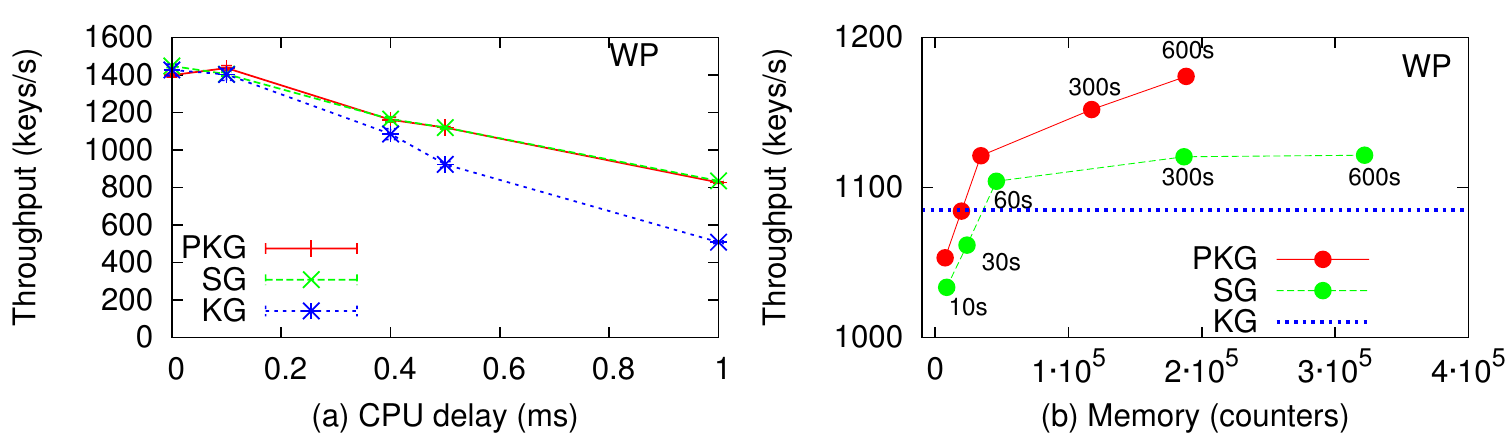}
	\includegraphics[scale=0.78]{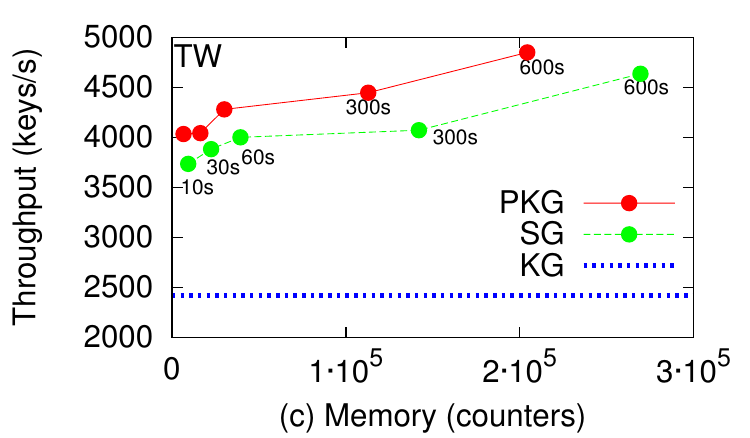}
	\caption{(a) Throughput for \pkgs, \sg and \kg for different CPU delays in WP. (b)-(c) Average throughput for \pkgs and \sg vs. average memory per worker for different aggregation periods in (b) WP (with CPU delay = 0.4ms) and (c) TW (with no CPU delay).}
	\label{fig:throughput_memory_cpu_aggr}
\end{center}
\vspace{-\baselineskip}
\end{figure*}

\spara{Q5.}
We implement and test \pkgs on Apache Storm, a popular distributed stream processing engine (\dspe).\footnote{\pkgs is integrated in the latest release (v0.10) of Storm.}
We perform an experiment by running a streaming top-k word count example and comparing \pkgs, \kg, and \sg on the TW dataset.
We chose word count as it is one of the simplest possible examples, thus limiting the number of confounding factors.
It is also representative of many data mining algorithms as the ones described in Section~\ref{sec:applications} (e.g., counting frequent items or co-occurrences of feature-class pairs).
Due to the requirement of real-world deployment on a \dspe, we ignore techniques that require coordination (i.e., \potc and On-Greedy).

For WP, we use a topology configuration with 8 sources, along with 10 workers for \kg, and 8 workers and 2 aggregators for \pkgs and \sg.
For TW, we use 16 workers for \kg, and 8 workers and 8 aggregators for \sg and \pkgs.
Both topologies run on a Storm cluster of 15 virtual servers.
The difference in the setup for TW is to compensate for the 10 times larger number of unique keys and 50 times larger number of messages processed in the system.
Note that in this experiment each message in the original stream generates $\approx$$10$-$15$ messages for the workers, one for each word contained in the tweet.
Therefore we compute throughput as number of words (i.e., keys) processed per second.
We report overall throughput, end-to-end latency, and memory usage.

In the first experiment we use WP and emulate different levels of CPU consumption per key by adding a fixed delay to the processing.
We prefer this solution over implementing a specific application in order to be able to control the load on the workers.
We choose a range that is able to bring our configuration to a saturation point, although the raw numbers would vary for different setups.
Even though real deployments rarely operate at saturation point, \pkgs allows better resource utilization, therefore supporting the same workload on a smaller number of machines.
In this case, the minimum delay ($0.1$ms) corresponds to reading approximately 400kB sequentially from memory, while the maximum one ($1$ms) to $\frac{1}{10}$-th of a disk seek.\footnote{\url{http://brenocon.com/dean_perf.html}}
Nevertheless, even more expensive tasks exist: parsing a sentence with NLP tools can take up to $500$ms.\footnote{\url{http://nlp.stanford.edu/software/parser-faq.shtml\#n}}

The system does not perform aggregation in this setup, as we are only interested in the raw effect on the workers.
Figure~\ref{fig:throughput_memory_cpu_aggr}(a) shows the throughput achieved when varying the CPU delay for the three partitioning strategies on WP.
Regardless of the delay, \sg and \pkgs perform similarly, and their throughput is higher than \kg.
The throughput of \kg is reduced by $\approx$$60\%$ when the CPU delay increases tenfold, while the impact on \pkgs and \sg is smaller ($\approx$$37\%$ decrease).
We deduce that reducing the imbalance is critical for clusters operating close to their saturation point, and that \pkgs is able to handle bottlenecks similarly to \sg and better than \kg.
In addition, the imbalance generated by \kg translates into longer latencies for the application, as shown in Table~\ref{tab:latencies}.
When the workers are heavily loaded, the average latency with \kg is up to $45\%$ larger than with \pkgs. 
Finally, the benefits of \pkgs over \sg regarding memory are substantial.
Overall, \pkgs ($3.6M$ counters) requires about $30\%$ more memory than \kg ($2.9M$ counters), but about half the memory of \sg ($7.2M$ counters).

\begin{table}[t]
\caption{Average latency per message for different partitioning schemes, CPU delays, and aggregation periods for WP.}
\centering
\small
\begin{tabular}{l c c c c c c}
\toprule
Scheme			&		\multicolumn{3}{c}{CPU delay D (ms)}				&	\multicolumn{3}{c}{Aggregation period T (s)}			\\
			\cmidrule(lr){2-4}	\cmidrule(lr){5-7}
			&	D=0.1		&	D=0.5		&	D=1			&	T=10			&	T=30			&	T=60			\\
\midrule
\pkgs		&	\num{3,81}	&	\num{6,24}	&	\num{11,01}	&	\num{6,93}	&	\num{6,79}	&	\num{6,47}	\\
\sg			&	\num{3,66}	&	\num{6,11}	&	\num{10,82}	&	\num{7,01}	&	\num{6,75}	&	\num{6,58}	\\
\kg			&	\num{3,65}	&	\num{9,82}	&	\num{19,35}	&				&				&				\\
\bottomrule
\end{tabular}
\label{tab:latencies}
\vspace{-\baselineskip}
\end{table}

In the second experiment, we fix the CPU delay to $0.4$ms per key, as it is the saturation point for \kg in our setup with WP.
We activate the aggregation of counters at different time intervals $T$ to emulate different application policies for when to receive up-to-date top-k word counts.
In this case, \pkgs and \sg need additional memory compared to \kg to keep partial counters.
Shorter aggregation periods reduce the memory requirements, as partial counters are flushed often, at the cost of a higher number of aggregation messages.
Figure~\ref{fig:throughput_memory_cpu_aggr}(b) shows the relationship between average throughput and memory overhead per worker for \pkgs and \sg in WP.
The throughput of \kg is shown for comparison.
For all values of aggregation period, \pkgs achieves higher throughput than \sg, with lower memory overhead and similar average latency per message. 
When the aggregation period is above $30$s, the benefits of \pkgs compensate its extra overhead and its overall throughput is higher than when using \kg.

In Figure~\ref{fig:throughput_memory_cpu_aggr}(c) we show the results on TW.
However, in this case we do not apply any artificial CPU delay, as the system naturally reaches a saturation point for \kg due to the 50 times larger load of messages processed.
The previous observations between \pkgs and \sg on WP for memory overhead vs. throughput are also confirmed on TW.
In particular, for an aggregation period of one minute, \pkgs improves the throughput nearly 175\% compared to \kg and reduces the memory overhead by almost 30\% compared to \sg . 
Interestingly, \kg fails to reach similar throughput levels as either \pkgs or \sg, due to the larger load on the workers assigned with the most frequent keys.
We anticipate these performance results to be representative of a real streaming application running on a \dspe deployed on a small storm cluster.
These results show that \pkg is a viable solution for realistic deployments that are challenging for other partitioning schemes.



\section{Related Work}
\label{sec:rel-work}
Various works in the literature either extend the theoretical results from the power of two choices, or apply them to the design of large-scale systems for data processing.

\spara{Theoretical results.}\label{sec:theor_choices}
Load balancing in a \dspe can be seen as a balls-and-bins problem, where $m$ balls are to be placed in $n$ bins.
The power of two choices has been extensively researched from a theoretical point of view for balancing the load among machines~\cite{mitzenmacher2001power,mitzenmacher2001potc-survey}. 
Previous results consider each ball equivalent.
For a \dspe, this assumption holds if we map balls to messages and bins to servers.
However, if we map balls to {\em keys}, more popular keys should be considered heavier.
\citet{talwar2007weightedcase} tackle the case where each ball has a weight drawn independently from a fixed weight distribution $\mathcal{X}$.
They prove that, as long as $\mathcal{X}$ is ``smooth'', the expected imbalance is independent of the number of balls. 
However, the solution assumes that $\mathcal{X}$ is known beforehand, which is not the case in a streaming setting.
Thus, in our work we take the standard approach of mapping balls to messages.


Another assumption common in previous works is that there is a single source of balls. 
Existing algorithms that extend \potc to multiple sources execute several rounds of intra-source coordination before taking a decision~\cite{adler1995parallelrandomized,lenzen2011parallelrandomized, park2011multiplechoices}.
These techniques incur a significant coordination overhead, which becomes prohibitive in a \dspe that handles thousands of messages per second.
 

\spara{Stream processing systems.}
Existing load balancing techniques for \dspes are analogous to key grouping with rebalancing~\citep{shah2003flux,cherniack2003scalable,xing2005dynamic,gedik2013partitioning,balkesen2013adaptive,castro2013integrating}.
In our work, we consider operators that allow replication and aggregation, similar to a standard combiner in map-reduce, and show that it is sufficient to balance load among two replicas based local load estimation.
We refer to Section~\ref{sec:existing-partitioning} for a more extensive discussion of key grouping with rebalancing.
Flux monitors the load of each operator, ranks servers by load, and migrates operators from the most loaded to the least loaded server, from the second most loaded to the second least loaded, and so on~\citep{shah2003flux}.
Aurora* and Medusa propose policies to migrating operators in \dspes and federated \dspes~\citep{cherniack2003scalable}.
Borealis uses a similar approach but it also aims at reducing the correlation of load spikes among operators placed on the same server~\citep{xing2005dynamic}.
This correlation is estimated by using a finite set of load samples taken in the recent past.
\citet{gedik2013partitioning} developed a partitioning function (a hybrid between explicit mapping and consistent hashing of items to servers) for stateful data parallelism in DSPEs that leverages item frequencies to control migration cost and imbalance in the system.
Similarly, \citet{ balkesen2013adaptive} proposed frequency-aware hash-based partitioning to achieve load balance.
\citet{castro2013integrating} propose integrating common operator state management techniques for both checkpointing and migration.

\spara{Other distributed systems.}
Several storage systems use consistent hashing to allocate data items to servers~\citep{karger1997consistent}.
Consistent hashing substantially produces a random allocation and is designed to deal with systems where the set of servers available varies over time.
In this paper, we propose replicating \dspe operators on two servers selected at random.
One could use consistent hashing also to select these two replicas, using the replication technique used by Chord~\citep{stoica2001chord} and other systems.

Sparrow~\cite{ousterhout2013sparrow} is a stateless distributed job scheduler that exploits a variant of the power of two choices~\citep{park2011multiplechoices}. 
It employs batch probing, along with late binding, to assign m tasks of a job to the least loaded of $d \times m$ randomly selected workers ($d \geq 1$). 
Sparrow considers only independent tasks that can be executed by any worker.
In \dspes, a message can only be sent to the workers that are accumulating the state corresponding to the key of that message.
Furthermore, \dspes deal with messages that arrive at a much higher rate than Sparrow's fine-grained tasks, so we prefer to use local load estimation.

In the domain of graph processing, several systems have been proposed to solve the load balancing problem, e.g., Mizan~\cite{khayyat2013mizan}, GPS~\cite{salihoglu2013gps}, and xDGP~\cite{vaqueroxdgp}.
Most of these systems perform dynamic load rebalancing at runtime via vertex migration.
Section~\ref{sec:preliminaries} already discusses why rebalancing is impractical in our context.
%

Finally, SkewTune~\cite{kwon2012skewtune} solves the problem of load balancing in MapReduce-like systems by identifying and redistributing the unprocessed data from the stragglers to other workers.
Techniques such as SkewTune are a good choice for batch processing systems, but cannot be directly applied to \dspes. 




\section{Conclusion}
Despite being a well-known problem in the literature, load balancing has not been exhaustively studied in the context of distributed stream processing engines.
Current solutions fail to provide satisfactory load balance when faced with skewed datasets. 
To solve this issue, we introduced \pkg, a new stream partitioning strategy that allows better load balance than key grouping while incurring less memory overhead than shuffle grouping.
Compared to key grouping, \pkgs is able to reduce the imbalance by up to several orders of magnitude, thus improving throughput and latency of an example application by up to 45\%. 
\pkgs has been integrated in Apache Storm release v0.10.

This work gives rise to further interesting research questions.
Is it possible to achieve good load balance without foregoing atomicity of processing of keys?
Is it feasible to design an algorithm that optimizes the trade-off between increase in memory usage and achieved load balance, by adapting to the characteristics of the input data stream?
And in a larger perspective, which other primitives can a \dspe offer to express algorithms effectively while making them run efficiently?
While most \dspes have settled on just a small set, the design space still remains largely unexplored.


%



\ifCLASSOPTIONcompsoc
  \section*{Acknowledgments}
\else
  \section*{Acknowledgment}
\fi
This work was produced during the internship of the first author at Yahoo Labs Barcelona.
The internship was supported by iSocial EU Marie Curie ITN project (FP7-PEOPLE-2012-ITN).

\ifCLASSOPTIONcaptionsoff
  \newpage
\fi



\bibliographystyle{IEEEtranNAT}
%
\bibliography{references}




%

\vspace{-1cm}

\begin{IEEEbiography}[{\includegraphics[width=1in, height=1.25in,clip,keepaspectratio]{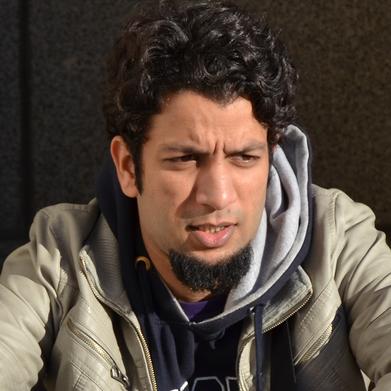}}]{Muhammad Anis Uddin Nasir}
is a PhD student at KTH Royal Institute of Technology, working under the Marie Curie Initial Training Network Project called iSocial. 
He finished a European Master in Distributed Computing from KTH Royal Institute of Technology and UPC, Polytechnic University of Catalunya. He holds a Bachelors in Computer Engineering from National University of Science and Technology, Pakistan.
More information can be found at \url{https://www.kth.se/profile/anisu/}.
\end{IEEEbiography}

\vspace{-1.3cm}

\begin{IEEEbiography}[{\includegraphics[width=1in, height=1.25in,clip,keepaspectratio]{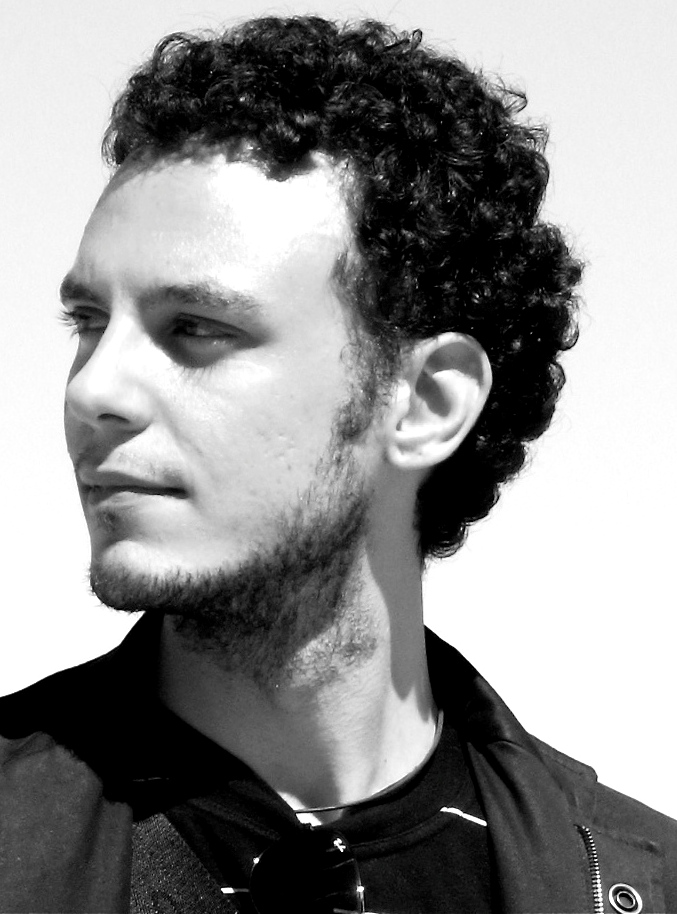}}]{Gianmarco De Francisci Morales}
is a Visiting Scientist at Aalto University, Helsinki.
He previously worked as a Research Scientist at Yahoo Labs in Barcelona.
His research focuses on scalable data mining, with a particular emphasis on Web mining and Data-Intensive Scalable Computing systems.
He is an active member of the Apache Software Foundation, working on the Hadoop ecosystem, and a committer for the Apache Pig project.
He is one of the lead developers of Apache SAMOA, an open-source platform for mining big data streams.
More information at \url{http://gdfm.me}.
\end{IEEEbiography}

\vspace{-1.3cm}

\begin{IEEEbiography}[{\includegraphics[width=1in, height=1.25in,clip,keepaspectratio]{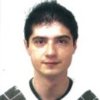}}]{David Garc\'ia-Soriano} 
is a Postdoctoral Researcher at Yahoo Labs Barcelona. He received his undergraduate degrees in Computer Science and Mathematics from the Complutense
University of
Madrid, and his PhD from the University of Amsterdam. 
His research interests include sublinear-time algorithms, learning, approximation algorithms, and large-scale problems in data mining and machine
learning.
More information at \url{https://sites.google.com/site/elhipercubo/}.
\end{IEEEbiography}

\vspace{-1.3cm}

\begin{IEEEbiography}[{\includegraphics[width=1in,height=1.25in,clip,keepaspectratio]{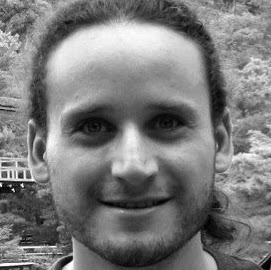}}]{Nicolas Kourtellis}
is a Postdoctoral Researcher in the Web Mining Research Group at Yahoo Labs Barcelona.
He received a PhD in Computer Science and Engineering from the University of South Florida in 2012.
He is interested in the network analysis of large-scale systems and social graphs, and property extraction useful in the design of improved socially-aware distributed systems.
More information at  \url{http://labs.yahoo.com/author/kourtell/}.
\end{IEEEbiography}

\vspace{-1.3cm}

\begin{IEEEbiography}[{\includegraphics[width=1in,height=1.25in,clip,keepaspectratio]{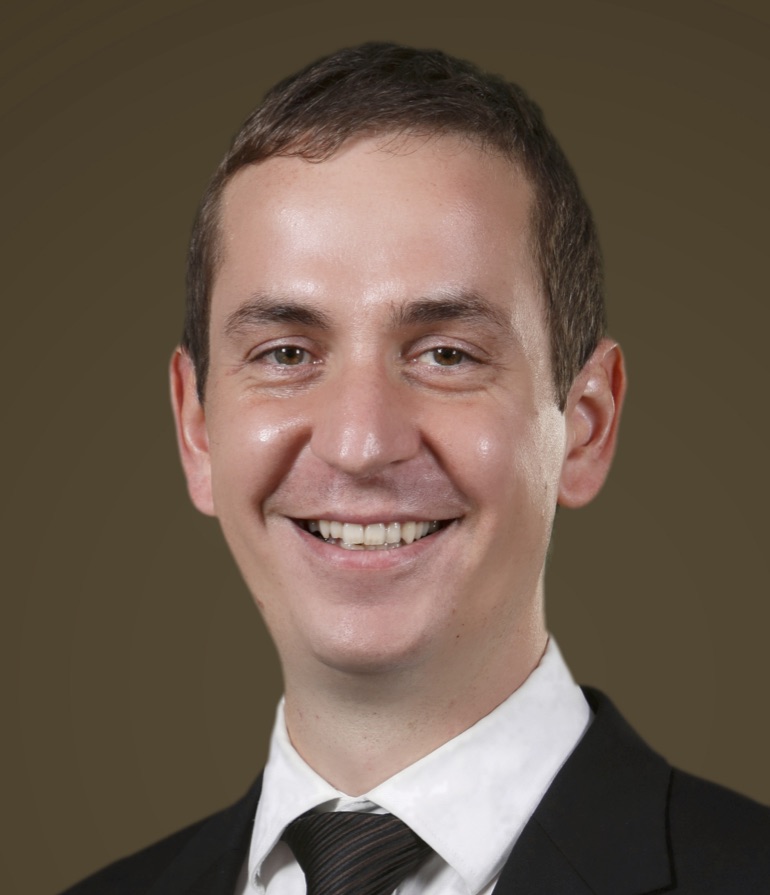}}]{Marco Serafini}
is a Scientist at Qatar Computing Research Institute, where he works on the scalability, dependability, and consistency of large-scale distributed systems and databases.
Before joining QCRI, he spent three years at Yahoo! Research Barcelona, working on Zookeeper, tolerance of data corruption and Arbitrary State Corruption faults, and social networking systems.
More information at \url{http://www.qcri.qa/our-people/marcoserafini}
\end{IEEEbiography}






\end{document}